\documentclass[12pt]{article}
\usepackage{a4wide}
\usepackage{latexsym}
\usepackage{amsmath}
\usepackage{amsfonts}
\usepackage{amscd}
\usepackage{cite}
\usepackage{placeins}

\usepackage{pslatex}
\usepackage{graphicx}
\usepackage[latin1,utf8]{inputenc}
\usepackage[T1]{fontenc}

\allowdisplaybreaks

\newcommand{\bq}{\begin{eqnarray}}
\newcommand{\eq}{\end{eqnarray}}
\newcommand{\eps}{\varepsilon}
\newcommand{\Eulerconstant}{\gamma_{\mathrm{E}}}

\newcommand{\NL}{N_L}

\begin{document}

\thispagestyle{empty}

\begin{flushright}
  MITP/23-070
\end{flushright}

\vspace{1.5cm}

\begin{center}
  {\Large\bf Box integrals with fermion bubbles for low-energy measurements of the weak mixing angle \\
  }
  \vspace{1cm}
  {\large Nico B\"ottcher, Niklas Schwanemann and Stefan Weinzierl \\
  \vspace{1cm}
      {\small \em PRISMA Cluster of Excellence, Institut f{\"u}r Physik, }\\
      {\small \em Johannes Gutenberg-Universit{\"a}t Mainz,}\\
      {\small \em D - 55099 Mainz, Germany}\\
  } 
\end{center}

\vspace{2cm}

\begin{abstract}\noindent
  {
The Moller experiment and the P2 experiment aim at measuring the weak mixing angle at low scales.
The Moller experiment uses $e^- e^- \rightarrow e^- e^-$-scattering, the P2 experiment uses $e^- N \rightarrow e^- N$-scattering.
In both cases, two-loop electroweak corrections have to be taken into account, and here in particular diagrams which give rise to large
logarithms.
In this paper we compute a set of two-loop electroweak Feynman integrals for point-like particles, 
which are obtained from a box integral by the insertion of a light fermion loop.
By rationalising all occurring square roots we show that these Feynman integrals can be expressed in terms of multiple polylogarithms.
We present the results in a form, which makes the large logarithms manifest. 
We provide highly efficient numerical evaluation routines for these integrals.
   }
\end{abstract}

\vspace*{\fill}

\newpage

\section{Introduction}
\label{sect:intro}

The Moller experiment \cite{Benesch:2014bas} at Jefferson Lab 
and the P2 experiment \cite{Becker:2018ggl} at the MESA accelerator at Mainz University 
will measure the weak mixing angle at low scales.
The process studied at the Moller experiment is electron-electron scattering, 
the process studied at the P2 experiment is electron-nucleon scattering.
In both cases, the experimental programs require theory input in the form of precision calculations.

Of particular importance are diagrams, which give rise to large logarithms.
Neglecting the electron mass, 
we have for both experiments the hierarchy of scales
\bq
 -t \;\; \ll \;\; s \sim m_N^2 \;\; \ll \;\; m_W^2 \sim m_Z^2,
\eq
where $s$ and $t$ are the Mandelstam variables, $m_N$ the nucleon mass (only relevant for the P2 experiment)
and $m_W$ and $m_Z$ the mass of the $W$-boson and $Z$-boson, respectively.
Large logarithms arise due to the smallness of $(-t)$ and/or due to the large mass of the heavy gauge bosons.
In this paper we consider a box diagram for point-like particles with the insertion of a light fermion loop both for the 
Moller experiment and the P2 experiment.
This diagram is shown in fig.~\ref{fig_box}.
The wavy lines in this diagram may either be photons (massless) or heavy gauge bosons.
In the case of M{\o}ller scattering the green line in fig.~\ref{fig_box} is massless,
for $e^- N \rightarrow e^- N$-scattering the green line in fig.~\ref{fig_box} is massive with mass $m_N$.
Depending on the mass configuration we have to calculate in total eight different cases, which we label
topology $A$ to topology $H$.
The case where all lines are massless is rather easy. 
This case will be labelled as topology $H$.
The opposite case, where all wavy lines have the mass of a heavy gauge boson (say $m_Z$) and the green line has the mass $m_N$
is rather involved and state-of-the-art in Feynman integral calculations.
The complication arises from four square roots, which are associated with this topology.
This latter case will be labelled as topology $A$.
We compute the master integrals for all topologies $A$-$H$.

As a side remark we note that crossing the diagrams with a massive green line gives diagrams relevant to two-loop electroweak corrections
and mixed two-loop QCD/electroweak corrections to the production of a pair of heavy particles at the LHC.

In this paper we perform an analytic calculation of these Feynman integrals.
We note that a subset of these integrals together with other related integrals 
have been computed previously with numerical methods \cite{Dubovyk:2016aqv,Du:2019evk,Erler:2022ckm,Dubovyk:2022frj,Armadillo:2022ugh}
or asymptotic approximations \cite{Aleksejevs:2015dba,Aleksejevs:2012xua}.

We compute the Feynman integrals with the help of the method of differential equations \cite{Kotikov:1990kg,Kotikov:1991pm,Gehrmann:1999as}:
Using integration-by-parts identities \cite{Chetyrkin:1981qh}
we first derive a differential equation for a pre-canonical basis of master integrals.
This differential equation is in general not in an $\eps$-factorised form.
We then construct a new basis, such that the differential equation is transformed to an $\eps$-factorised form \cite{Henn:2013pwa}.
In doing so, we unavoidably need to introduce square roots.
In total we encounter five different square roots.
It is advisable to treat each topology separately, as not all five square roots occur simultaneously in a given topology.
The most complicated case is topology $A$ with four occurring square roots.
For every topology we may simultaneously rationalise all occurring square roots.
This shows that the Feynman integrals can be expressed in terms of multiple polylogarithms.
Furthermore, by choosing an appropriate boundary point and by isolating trailing zeros in powers of single logarithms we may
make all large logarithms manifest.
For all topologies we provide a highly efficient numerical {\tt C++}-program,
which evaluates the master integrals in the kinematic region of interest
with arbitrary precision.  

For phenomenological applications our results can be used as follows: 
Let us first consider the case of M{\o}ller scattering.
It is well-known that all occurring Feynman integrals in a scattering amplitude
can be reduced to master integrals.
Standard integration-by-parts reduction programs like
{\tt FIRE/Litered} \cite{Smirnov:2008iw,Smirnov:2019qkx,Lee:2012cn,Lee:2013mka},
{\tt Reduze} \cite{Studerus:2009ye,vonManteuffel:2012np},
{\tt Kira} \cite{Maierhoefer:2017hyi,Klappert:2020nbg} or
{\tt FiniteFlow} \cite{Peraro:2016wsq,Peraro:2019svx}
facilitate this task.
In general, these programs reduce Feynman integrals to a pre-canonical basis of master integrals. 
Our starting point is the default choice of {\tt Kira} for a pre-canonical basis of master integrals.
The integration-by-parts reduction programs can be used to convert this basis to any other pre-canonical basis. 
The pre-canonical basis is related by a rotation matrix to the basis of uniform transcendental weight constructed in this paper.
The rotation matrix and its inverse are given in the supplementary electronic file
attached to the arxiv version of this article.
The master integrals of uniform transcendental weight are computed with the provided {\tt C++}-programs,
again given in the supplementary electronic file attached to the arxiv version of this article.

The case of electron-nucleon scattering has additional complications due to the hadronic nature of the nucleon.
For the details how calculations are done in this situation we refer to refs.~\cite{Marciano:1982mm,Marciano:1983ss,Erler:2003yk,Zhou:2009nf,Rislow:2010vi,Blunden:2011rd,Hall:2015loa,Gorchtein:2016qtl,Erler:2019rmr}.
The main interest is the coefficient of the large logarithm related to the heavy boson mass and we provide the master integrals to extract this coefficient.
Let us emphasise that although our calculation is with point-like particles, 
the inclusion of form factors for the coupling of a nucleon to a gauge boson is straightforward, as long as the form factors are modelled by rational functions
in the momenta, which do not introduce new singularities.
On the other hand, the inclusion of a nucleon resonance introduces another kinematic variable
and leads to loop integrals beyond the ones considered in this paper.

This paper is organised as follows:
In section~\ref{sect:preliminaries} we show that due to partial fractioning we only need to compute a reduced graph.
In section~\ref{sect:notation} we introduce the notation for the Feynman integrals.
In section~\ref{sect:master_integrals} we present for all topologies a basis of master integrals of uniform transcendental weight. 
In section~\ref{sect:differential_equations} we give the differential equation for each topology and list
all differential one-forms appearing in the differential equations.
The differential one-forms are dlog-forms with algebraic functions as argument.
The algebraic part is given by five square roots. In total we encounter five different square roots.
In section~\ref{sect:rationalisation} we show that for each topology all occurring square roots can be rationalised.
As a consequence, all Feynman integrals can be expressed in terms of multiple polylogarithms.
In order to solve the differential equations, we need boundary values. 
These are given in section~\ref{sect:boundary}.
In section~\ref{sect:results} we present numerical results and the leading large logarithms.
Finally, our conclusions are given in section~\ref{sect:conclusions}.
In appendix~\ref{sect_master_sectors} we show for all master integrals the corresponding diagrams.
In appendix~\ref{sect:supplement} we describe the content of the supplementary electronic file
attached to the arxiv version of this article.
In appendix~\ref{sect_one_loop} we collect for convenience 
the corresponding one-loop integrals in the same notation as used for the two-loop integrals.


\section{Preliminaries}
\label{sect:preliminaries}

\begin{figure}
\begin{center}
\includegraphics[scale=1.0]{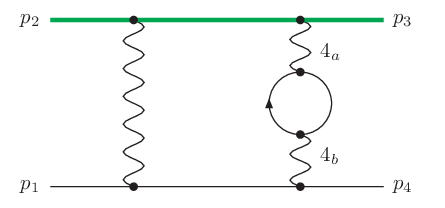}
\end{center}
\caption{
The two-loop Feynman graph. Wavy lines are either photons (massless) or heavy gauge bosons (massive).
The mass of the green line is either zero (this case corresponds to the Moller experiment) or massive with mass $m_N$ 
(this case corresponds to the P2 experiment).
All other particles are assumed to be massless.
}
\label{fig_box}
\end{figure}

We are interested in the Feynman graph shown in fig.~\ref{fig_box}.
The green line could either be a massless fermion or a nucleon (with non-zero mass $m_N$). 
The wavy lines are either massless gauge bosons (photons) or massive gauge bosons ($Z$-bosons or $W$-bosons).
We are in particular interested in the case, where at least one of the wavy lines is a massive gauge boson.
The case where all of them are photons is significantly simpler and only included for completeness.
The black solid lines correspond to massless fermions.
For $e^- e^- \rightarrow e^- e^-$-scattering we take the green line to be massless, for $e^- N \rightarrow e^- N$-scattering
we take the green line to be massive.
It is sufficient to focus on the case, where the heavy gauge boson is the $Z$-boson.
As we neglect the electron mass, electrons and neutrinos both have zero mass. Furthermore we do not distinguish
between the proton and the neutron mass.
With these approximations the case with heavy $W$-bosons gives rise to exactly the same integrals. 
In the following we will denote the heavy gauge boson mass by $m_Z$.

We recall that any tensor integral can be reduced to scalar integrals \cite{Tarasov:1996br,Tarasov:1997kx}.
Therefore, although we are interested in integrals with fermions and gauge bosons, what we have to calculate are scalar integrals.

There is one immediate simplification:
The momenta flowing through the propagators labelled $4_a$ and $4_b$ in fig.~\ref{fig_box} are the same.
The two propagators may have equal mass (either $m_{4_a}=m_{4_b}=0$ or $m_{4_a}=m_{4_b}=m_Z$)
or unequal mass (either $m_{4_a}=0, m_{4_b}=m_Z$ or $m_{4_a}=m_Z, m_{4_b}=0$).
In the latter case we may use partial fraction decomposition
\bq
 \frac{1}{\left(-q^2+m_{4_a}^2\right)\left(-q^2+m_{4_b}^2\right)}
 & = &
 \frac{1}{m_{4_b}^2-m_{4_a}^2} \left[ \frac{1}{\left(-q^2+m_{4_a}^2\right)} - \frac{1}{\left(-q^2+m_{4_b}^2\right)} \right].
\eq
\begin{figure}
\begin{center}
\includegraphics[scale=1.0]{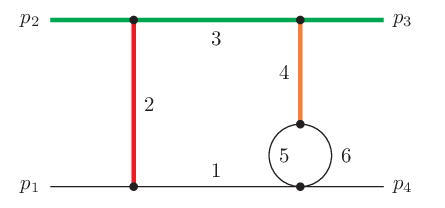}
\end{center}
\caption{
The reduced Feynman graph.
}
\label{fig_reduced_box}
\end{figure}
We therefore have to consider only
the reduced topology shown in fig.~\ref{fig_reduced_box}
with
\bq
 m_2, m_4 \; \in \; \left\{ 0, m_Z \right\},
 & &
 m_3 \; \in \; \left\{ 0, m_N \right\}.
\eq
The external momenta satisfy
\bq
 p_1^2
 \; = \;
 p_4^2
 \; = \;
 0,
 & &
 p_2^2
 \; = \;
 p_3^2
 \; = \;
 m_3^2.
\eq
We denote the Mandelstam variables by
\bq
 s \; = \; \left(p_1+p_2\right)^2,
 \;\;\;\;\;\;
 t \; = \; \left(p_2+p_3\right)^2,
 \;\;\;\;\;\;
 u \; = \; \left(p_1+p_3\right)^2.
\eq


\section{Notation}
\label{sect:notation}

\begin{figure}
\begin{center}
\includegraphics[scale=1.0]{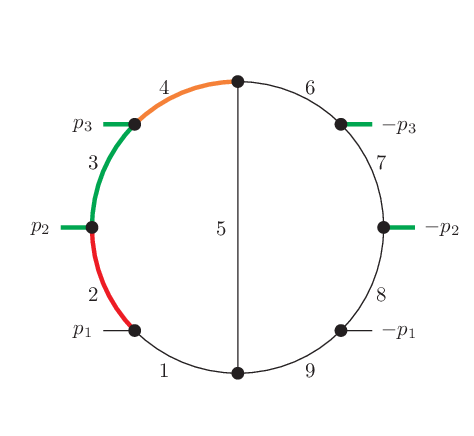}
\end{center}
\caption{
The auxiliary graph.
Green lines correspond to a particle with mass $m_3$, 
orange lines correspond to a particle with mass $m_4$ and
red lines correspond to a particle with mass $m_2$.
}
\label{fig_auxiliary-graph}
\end{figure}

We need to consider an auxiliary graph associated to the reduced graph of fig.~\ref{fig_reduced_box}, 
such that any scalar product involving at least one loop momentum can be expressed as a linear combination of inverse
propagators. With three independent external momenta and two independent loop momenta this associated graph
must have nine internal propagators.
We therefore consider the family of integrals
\bq
\label{def_integral}
 I_{\nu_1 \nu_2 \nu_3 \nu_4 \nu_5 \nu_6 \nu_7 \nu_8 \nu_9}
 & = &
 e^{2 \Eulerconstant \eps}
 \left(\mu^2\right)^{\nu-D}
 \int \frac{d^Dk_1}{i \pi^{\frac{D}{2}}} \frac{d^Dk_2}{i \pi^{\frac{D}{2}}}
 \prod\limits_{j=1}^9 \frac{1}{ P_j^{\nu_j} },
\eq
where $\Eulerconstant$ denotes the Euler-Mascheroni constant, $D=4-2\eps$ is the number of space-time dimensions,
$\mu$ is an arbitrary scale introduced to render the Feynman integral dimensionless and
the quantity $\nu$ is given by
\bq
 \nu & = &
 \sum\limits_{j=1}^9 \nu_j.
\eq
We will further use the notation $p_{ij}=p_i+p_j$, $p_{ijk}=p_i+p_j+p_k$.
The inverse propagators are given by
\begin{align}
 P_1 & = -k_1^2,
 &
 P_2 & = -\left(k_1-p_1\right)^2 + m_2^2,
 &
 P_3 & = -\left(k_1-p_{12}\right)^2 + m_3^2,
 \nonumber \\
 P_4 & = -\left(k_1-p_{123}\right)^2 + m_4^2,
 &
 P_5 & = -k_2^2,
 &
 P_6 & = -\left(k_1+k_2-p_{123}\right)^2,
 \nonumber \\
 P_7 & = -\left(k_1+k_2+p_{12}\right)^2,
 & 
 P_8 & = -\left(k_1+k_2-p_{1}\right)^2,
 &
 P_9 & = -\left(k_1+k_2\right)^2.
\end{align}
The graph for this family is shown in fig.~\ref{fig_auxiliary-graph}.
We are interested in the sectors, for which we have
\bq
 \nu_7, \nu_8, \nu_9  & \le & 0.
\eq
We define a sector id (or topology id) by
\bq
\label{def_sector_id}
 \mathrm{id}
 & = & \sum\limits_{j=1}^9 2^{j-1} \Theta\left(\nu_j-\frac{1}{2}\right).
\eq
Here, $\Theta(x)$ denotes the Heaviside step function.
Since we assume $\nu_j \in {\mathbb Z}$, the shift by $(-1/2)$ avoids any ambiguity in the definition of $\Theta(0)$.

We have to consider eight cases, depending on whether the masses $(m_2,m_3,m_4)$ are non-zero or zero.
We define eight topologies
\begin{align}
 A: \;\; & \left(m_2,m_3,m_4\right) = \left(m_Z,m_N,m_Z\right),
 &
 E: \;\;& \left(m_2,m_3,m_4\right) = \left(m_Z,0,m_Z\right),
 \nonumber \\
 B: \;\;& \left(m_2,m_3,m_4\right) = \left(0,m_N,m_Z\right),
 &
 F: \;\;& \left(m_2,m_3,m_4\right) = \left(0,0,m_Z\right),
 \nonumber \\
 C: \;\;& \left(m_2,m_3,m_4\right) = \left(m_Z,m_N,0\right),
 &
 G: \;\;& \left(m_2,m_3,m_4\right) = \left(m_Z,0,0\right),
 \nonumber \\
 D: \;\;& \left(m_2,m_3,m_4\right) = \left(0,m_N,0\right),
 &
 H: \;\;& \left(m_2,m_3,m_4\right) = \left(0,0,0\right).
\end{align}
We write
\bq
 I_{\nu_1 \nu_2 \nu_3 \nu_4 \nu_5 \nu_6 \nu_7 \nu_8 \nu_9}^X,
 & &
 X \; \in \; \left\{ A,B,C,D,E,F,G,H \right\}
\eq
to denote the corresponding integrals.

The total number of master integrals for the various topologies are
shown in table~\ref{table_masters}.
\begin{table}
\begin{center}
\begin{tabular}{|c|c||c|c|}
\hline
 Topology & master integrals & topology & master integrals \\
\hline
 $A$ & $15$ & $E$ & $10$ \\
 $B$ & $11$ & $F$ & $7$ \\
 $C$ & $8$ & $G$ & $5$ \\
 $D$ & $5$ & $H$ & $3$ \\
\hline
\end{tabular}
\end{center}
\caption{
The number of master integrals for a given topology.
}
\label{table_masters}
\end{table}
In defining master integrals of uniform transcendental weight
we will encounter five square roots. 
\begin{table}
\begin{center}
\begin{tabular}{|c|c||c|c|}
\hline
 Topology & roots & topology & roots \\
\hline
 $A$ & $r_1, r_2, r_3, r_5$ & $E$ & $r_2, r_4$ \\
 $B$ & $r_1, r_3$ & $F$ & $-$ \\
 $C$ & $r_1, r_3$ & $G$ & $-$ \\
 $D$ & $r_1$ & $H$ & $-$ \\
\hline
\end{tabular}
\end{center}
\caption{
The various roots appearing in a given topology.
}
\label{table_roots}
\end{table}
These are given by
\bq
\label{def_square_roots}
 r_1 & = & \sqrt{-t\left(4m_N^2-t\right)},
 \nonumber \\
 r_2 & = & \sqrt{-t\left(4m_Z^2-t\right)},
 \nonumber \\
 r_3 & = & \sqrt{-m_Z^2\left(4m_N^2-m_Z^2\right)},
 \nonumber \\
 r_4 & = &
 \sqrt{s t \left[s t -4m_Z^2\left(m_Z^2 + s\right)\right]}.
 \nonumber \\
 r_5 & = &
 \sqrt{-t\left[-t \left(m_N^2-s\right)^2+4m_Z^2\left(m_Z^2 s + \left(m_N^2-s\right)^2\right)\right]}.
\eq
We have chosen the arguments of the five square roots such that in the region of interest 
($t<0$, $s>0$, $m_Z^2 \gg m_N^2,s,(-t)$)
the arguments of all five roots are positive.
In this region we chose the sign of the square roots such that all five roots are positive.
Table~\ref{table_roots} shows for each topology the roots appearing in this topology.

The first graph polynomial \cite{Bogner:2010kv} for the auxiliary graph shown in fig.~(\ref{fig_auxiliary-graph}) reads
\bq
\lefteqn{
 {\mathcal U}\left(a_1,a_2,a_3,a_4,a_5,a_6,a_7,a_8,a_9\right)
 = } & & \\ 
 & &
 \left(a_1+a_2+a_3+a_4\right)\left(a_6+a_7+a_8+a_9\right)
 + a_5 \left(a_1+a_2+a_3+a_4+a_6+a_7+a_8+a_9\right).
 \nonumber
\eq
We introduce an operator ${\bf i}^+$, 
which raises the power of the propagator $i$ by one and multiplies by $\nu_i$, e.g.
\bq
 {\bf 1}^+ I_{\nu_1 \nu_2 \nu_3 \nu_4 \nu_5 \nu_6 \nu_7 \nu_8 \nu_9} & = &
 \nu_1 \cdot I_{(\nu_1+1) \nu_2 \nu_3 \nu_4 \nu_5 \nu_6 \nu_7 \nu_8 \nu_9}.
\eq
The notation with an extra prefactor $\nu_j$ follows ref.~\cite{Weinzierl:2022eaz}.
In addition we define the operator ${\bf D}^-$,
which lowers the dimension of space-time by two units through
\bq
 {\bf D}^- I_{\nu_1 \nu_2 \nu_3 \nu_4 \nu_5 \nu_6 \nu_7 \nu_8 \nu_9}\left( D \right)
 & = &
 I_{\nu_1 \nu_2 \nu_3 \nu_4 \nu_5 \nu_6 \nu_7 \nu_8 \nu_9}\left( D-2 \right).
\eq
The dimensional shift relations read \cite{Tarasov:1996br,Tarasov:1997kx}
\bq
\label{dim_shift_eq}
 {\bf D}^- I_{\nu_1 \nu_2 \nu_3 \nu_4 \nu_5 \nu_6 \nu_7 \nu_8 \nu_9}\left(D\right)
 =
 {\mathcal U}\left( {\bf 1}^+, {\bf 2}^+, {\bf 3}^+, {\bf 4}^+, {\bf 5}^+ , {\bf 6}^+, {\bf 7}^+ , {\bf 8}^+ , {\bf 9}^+ \right)
 I_{\nu_1 \nu_2 \nu_3 \nu_4 \nu_5 \nu_6 \nu_7 \nu_8 \nu_9}\left(D\right).
\eq


\section{The master integrals}
\label{sect:master_integrals}

We will treat each topology separately.
The main motivation is that this allows us to rationalise for each topology all occurring square roots.
This introduces a small redundancy, as the same sub-sectors may occur in more than one topology.
This redundancy provides an additional cross-check, as the sub-sectors are computed with different
rationalisations and different integration paths.

\subsection{Pre-canonical master integrals}

Standard integration-by-parts reduction programs like
{\tt FIRE/Litered} \cite{Smirnov:2008iw,Smirnov:2019qkx,Lee:2012cn,Lee:2013mka},
{\tt Reduze} \cite{Studerus:2009ye,vonManteuffel:2012np},
{\tt Kira} \cite{Maierhoefer:2017hyi,Klappert:2020nbg} or
{\tt FiniteFlow} \cite{Peraro:2016wsq,Peraro:2019svx}
are capable to express any relevant scalar Feynman integral as a linear combination of master integrals.
The chosen master integrals depend on the ordering criteria in the Laporta algorithm \cite{Laporta:2001dd}.
In general, the chosen master integrals are not of uniform weight.
Possible pre-canonical bases are:
\bq
 I^A & = &
 \left(
  I^A_{010011000},
  I^A_{\left(-1\right)10011000},
  I^A_{001011000},
  I^A_{000111000},
  I^A_{101011000},
  I^A_{011011000},
  I^A_{\left(-1\right)11011000},
 \right. \nonumber \\
 & & \left.
  I^A_{010111000},
  I^A_{001111000},
  I^A_{111011000},
  I^A_{111\left(-1\right)11000},
  I^A_{110111000},
  I^A_{101111000},
  I^A_{011111000},
  I^A_{111111000}
 \right)^T,
 \nonumber \\
 I^B & = &
 \left(
  I^B_{010011000},
  I^B_{001011000},
  I^B_{000111000},
  I^B_{101011000},
  I^B_{011011000},
  I^B_{010111000},
  I^B_{001111000},
  I^B_{111011000},
 \right. \nonumber \\
 & & \left.
  I^B_{101111000},
  I^B_{011111000},
  I^B_{111111000}
 \right)^T,
 \nonumber \\
 I^C & = &
 \left(
  I^C_{010011000},
  I^C_{\left(-1\right)10011000},
  I^C_{001011000},
  I^C_{101011000},
  I^C_{011011000},
  I^C_{\left(-1\right)11011000},
  I^C_{111011000},
 \right. \nonumber \\
 & & \left.
  I^C_{111\left(-1\right)11000}
 \right)^T,
 \nonumber \\
 I^D & = &
 \left(
  I^D_{010011000},
  I^D_{001011000},
  I^D_{101011000},
  I^D_{011011000},
  I^D_{111011000}
 \right)^T,
 \nonumber \\
 I^E & = &
 \left(
  I^E_{010011000},
  I^E_{\left(-1\right)10011000},
  I^E_{000111000},
  I^E_{101011000},
  I^E_{010111000},
  I^E_{111011000},
  I^E_{111\left(-1\right)11000},
 \right. \nonumber \\
 & & \left.
  I^E_{110111000},
  I^E_{101111000},
  I^E_{111111000}
 \right)^T,
 \nonumber \\
 I^F & = &
 \left(
  I^F_{010011000},
  I^F_{000111000},
  I^F_{101011000},
  I^F_{010111000},
  I^F_{111011000},
  I^F_{101111000},
  I^F_{111111000}
 \right)^T,
 \nonumber \\
 I^G & = &
 \left(
  I^G_{010011000},
  I^G_{\left(-1\right)10011000},
  I^G_{101011000},
  I^G_{111011000},
  I^G_{111\left(-1\right)11000}
 \right)^T,
 \nonumber \\
 I^H & = &
 \left(
  I^H_{010011000},
  I^H_{101011000},
  I^H_{111011000}
 \right)^T.
\eq
In the following, we will take these pre-canonical bases as our starting point.
Diagrams for all master sectors are shown in appendix~\ref{sect_master_sectors}.

\subsection{Master integrals of uniform transcendental weight}
\label{sect:master_integrals_uniform_transcendental_weight}

Below we present for all eight topologies master integrals of uniform transcendental weight.
They are related to the pre-canonical basis by
\bq
\label{def_fibre_transformation}
 J^X & = & U^X I^X, 
 \;\;\;\;\;\;
 X \in \left\{ A,B,C,D,E,F,G,H \right\}.
\eq
The dimension of the matrix $U^X$ is given by the number of master integrals for topology $X$.
The matrices $U^X$ are given in an electronic file
attached to the arxiv version of this article.
The master integrals of uniform transcendental weight are constructed by analysing the maximal cut
in the loop-by-loop Baikov representation \cite{Frellesvig:2017aai,Weinzierl:2022eaz}.
We have chosen the master integrals such that they simplify in 
kinematic limits (e.g. $m_N\rightarrow 0$ or $m_Z\rightarrow 0$) to the master integrals 
in the simpler topologies.

\subsubsection{Topology A}

A possible choice of master integrals of uniform transcendental weight for topology $A$ is given by:
\begin{alignat}{2}
 \mbox{Sector 50:} \;\;\;\; &
 J^{A}_{1}
 & = \;\; & 
 \eps^2 
 \left(\frac{m_Z^2-t}{\mu^2}\right)
 \; {\bf D}^- I^A_{010011000},
 \nonumber \\
 &
 J^{A}_{2}
 & = \;\; & 
 \eps^2 
 \; {\bf D}^- I^A_{010\left(-1\right)11000}
 -
 \eps^2 
 \left(\frac{m_Z^2}{\mu^2}\right)
 \; {\bf D}^- I^A_{010011000},
 \nonumber \\
 \mbox{Sector 52:} \;\;\;\; &
 J^{A}_{3}
 & = \;\; & 
 \eps \left(1+4\eps\right) 
 \left( \frac{m_N^2}{\mu^2} \right)
 \; {\bf D}^- I^A_{001011000},
 \nonumber \\
 \mbox{Sector 56:} \;\;\;\; &
 J^{A}_{4}
 & = \;\; & 
 \eps^2 
 \left(\frac{m_Z^2}{\mu^2}\right)
 \; {\bf D}^- I^A_{000111000},
 \nonumber \\
 \mbox{Sector 53:} \;\;\;\; &
 J^{A}_{5}
 & = \;\; & 
 -4 \eps^3 
 \left(\frac{m_N^2-s}{\mu^2}\right)
 I^A_{101012000},
 \nonumber \\
 \mbox{Sector 54:} \;\;\;\; &
 J^{A}_{6}
 & = \;\; & 
 \eps^3 
 \left(\frac{r_1}{\mu^2}\right)
 I^A_{011012000},
 \nonumber \\
 &
 J^{A}_{7}
 & = \;\; & 
 \eps^2 
 \left(\frac{r_3}{\mu^2}\right)
 \; {\bf D}^- I^A_{011\left(-1\right)11000}
 -
 \eps^2 
 \left(\frac{r_3}{\mu^2}\right) \left(\frac{m_Z^2}{\mu^2}\right)
 \; {\bf D}^- I^A_{011011000},
 \nonumber \\
 \mbox{Sector 58:} \;\;\;\; &
 J^{A}_{8}
 & = \;\; & 
 \eps^2 
 \left(\frac{r_2}{\mu^2}\right) \left(\frac{m_Z^2}{\mu^2}\right)
 \; {\bf D}^- I^A_{010111000}
 -
 \eps^2 
 \left(\frac{r_2}{\mu^2}\right)
 \; {\bf D}^- I^A_{010011000},
 \nonumber \\
 \mbox{Sector 60:} \;\;\;\; &
 J^{A}_{9}
 & = \;\; & 
 \eps^2 
 \left(\frac{m_Z^2}{\mu^2}\right)\left(\frac{r_3}{\mu^2}\right)
 \; {\bf D}^- I^A_{001111000}
 - \eps^2 
 \left(\frac{r_3}{\mu^2}\right)
 \; {\bf D}^- I^A_{001011000},
 \nonumber \\
 \mbox{Sector 55:} \;\;\;\; &
 J^{A}_{10}
 & = \;\; & 
 \eps^3
 \left(\frac{m_N^2-s}{\mu^2}\right) \left(\frac{m_Z^2-t}{\mu^2}\right)
 I^A_{111012000},
 \nonumber \\
 &
 J^{A}_{11}
 & = \;\; & 
 \eps^3
 \left(\frac{m_N^2-s}{\mu^2}\right)
 I^A_{111\left(-1\right)12000}
 -
 \eps^3
 \left(\frac{m_N^2-s}{\mu^2}\right) \left(\frac{m_Z^2}{\mu^2}\right)
 I^A_{111012000},
 \nonumber \\
 \mbox{Sector 59:} \;\;\;\; &
 J^{A}_{12}
 & = \;\; & 
 \eps^3
 \left(\frac{-t}{\mu^2}\right) \left(\frac{m_Z^2}{\mu^2}\right)
 I^A_{110112000},
 \nonumber \\
 \mbox{Sector 61:} \;\;\;\; &
 J^{A}_{13}
 & = \;\; & 
 \eps^3
 \left(\frac{m_N^2-s}{\mu^2}\right) \left(\frac{m_Z^2}{\mu^2}\right)
 I^A_{101112000},
 \nonumber \\
 \mbox{Sector 62:} \;\;\;\; &
 J^{A}_{14}
 & = \;\; & 
 \eps^3
 \left(\frac{r_1}{\mu^2}\right) \left(\frac{m_Z^2}{\mu^2}\right)
 I^A_{011112000},
 \nonumber \\
 \mbox{Sector 63:} \;\;\;\; &
 J^{A}_{15}
 & = \;\; & 
 \eps^3
 \left(\frac{r_5}{\mu^4}\right) \left(\frac{m_Z^2}{\mu^2}\right)
 I^A_{111112000}
 -
 \eps^3
 \left(\frac{r_5}{\mu^4}\right)
 I^A_{111012000}.
\end{alignat}

\subsubsection{Topology B}

A possible choice of master integrals of uniform transcendental weight for topology $B$ is given by:
\begin{alignat}{2}
 \mbox{Sector 50:} \;\;\;\; &
 J^{B}_{1}
 & = \;\; & 
 \eps^2 
 \left(\frac{-t}{\mu^2}\right)
 I^B_{010022000},
 \nonumber \\
 \mbox{Sector 52:} \;\;\;\; &
 J^{B}_{2}
 & = \;\; & 
 \eps \left(1+4\eps\right) 
 \left( \frac{m_N^2}{\mu^2} \right)
 \; {\bf D}^- I^B_{001011000},
 \nonumber \\
 \mbox{Sector 56:} \;\;\;\; &
 J^{B}_{3}
 & = \;\; & 
 \eps^2 
 \left(\frac{m_Z^2}{\mu^2}\right)
 \; {\bf D}^- I^B_{000111000},
 \nonumber \\
 \mbox{Sector 53:} \;\;\;\; &
 J^{B}_{4}
 & = \;\; & 
 -4 \eps^3 
 \left(\frac{m_N^2-s}{\mu^2}\right)
 I^B_{101012000},
 \nonumber \\
 \mbox{Sector 54:} \;\;\;\; &
 J^{B}_{5}
 & = \;\; & 
 \eps^3 
 \left(\frac{r_1}{\mu^2}\right)
 I^B_{011012000},
 \nonumber \\
 \mbox{Sector 58:} \;\;\;\; &
 J^{B}_{6}
 & = \;\; & 
 \eps^2 
 \left(\frac{m_Z^2}{\mu^2}\right)\left(\frac{m_Z^2-t}{\mu^2}\right)
 \; {\bf D}^- I^B_{010111000}
 - \eps^2 
 \left(\frac{m_Z^2}{\mu^2}\right)
 \; {\bf D}^- I^B_{010011000},
 \nonumber \\
 \mbox{Sector 60:} \;\;\;\; &
 J^{B}_{7}
 & = \;\; & 
 \eps^2 
 \left(\frac{m_Z^2}{\mu^2}\right)\left(\frac{r_3}{\mu^2}\right)
 \; {\bf D}^- I^B_{001111000}
 - \eps^2 
 \left(\frac{r_3}{\mu^2}\right)
 \; {\bf D}^- I^B_{001011000},
 \nonumber \\
 \mbox{Sector 55:} \;\;\;\; &
 J^{B}_{8}
 & = \;\; & 
 \eps^3
 \left(\frac{m_N^2-s}{\mu^2}\right) \left(\frac{-t}{\mu^2}\right)
 I^B_{111012000},
 \nonumber \\
 \mbox{Sector 61:} \;\;\;\; &
 J^{B}_{9}
 & = \;\; & 
 \eps^3
 \left(\frac{m_N^2-s}{\mu^2}\right) \left(\frac{m_Z^2}{\mu^2}\right)
 I^B_{101112000},
 \nonumber \\
 \mbox{Sector 62:} \;\;\;\; &
 J^{B}_{10}
 & = \;\; & 
 \eps^3
 \left(\frac{r_1}{\mu^2}\right) \left(\frac{m_Z^2}{\mu^2}\right)
 I^B_{011112000},
 \nonumber \\
 \mbox{Sector 63:} \;\;\;\; &
 J^{B}_{11}
 & = \;\; & 
 \eps^3
 \left(\frac{m_N^2-s}{\mu^2}\right) \left(\frac{m_Z^2-t}{\mu^2}\right) \left(\frac{m_Z^2}{\mu^2}\right)
 I^B_{111112000}
 -
 \eps^3
 \left(\frac{m_N^2-s}{\mu^2}\right) \left(\frac{m_Z^2}{\mu^2}\right)
 I^B_{111012000}.
\end{alignat}

\subsubsection{Topology C}

A possible choice of master integrals of uniform transcendental weight for topology $C$ is given by:
\begin{alignat}{2}
 \mbox{Sector 50:} \;\;\;\; &
 J^{C}_{1}
 & = \;\; & 
 \eps^2 
 \left(\frac{m_Z^2-t}{\mu^2}\right)
 \; {\bf D}^- I^C_{010011000},
 \nonumber \\
 &
 J^{C}_{2}
 & = \;\; & 
 \eps^2 
 \; {\bf D}^- I^C_{010\left(-1\right)11000},
 \nonumber \\
 \mbox{Sector 52:} \;\;\;\; &
 J^{C}_{3}
 & = \;\; & 
 \eps \left(1+4\eps\right) 
 \left( \frac{m_N^2}{\mu^2} \right)
 \; {\bf D}^- I^C_{001011000},
 \nonumber \\
 \mbox{Sector 53:} \;\;\;\; &
 J^{C}_{4}
 & = \;\; & 
 -4 \eps^3 
 \left(\frac{m_N^2-s}{\mu^2}\right)
 I^C_{101012000},
 \nonumber \\
 \mbox{Sector 54:} \;\;\;\; &
 J^{C}_{5}
 & = \;\; & 
 \eps^3 
 \left(\frac{r_1}{\mu^2}\right)
 I^C_{011012000},
 \nonumber \\
 &
 J^{C}_{6}
 & = \;\; & 
 \eps^2
 \left(\frac{r_3}{\mu^2}\right)
 \; {\bf D}^- I^C_{011\left(-1\right)11000},
 \nonumber \\
 \mbox{Sector 55:} \;\;\;\; &
 J^{C}_{7}
 & = \;\; & 
 \eps^3
 \left(\frac{m_N^2-s}{\mu^2}\right) \left(\frac{m_Z^2-t}{\mu^2}\right)
 I^C_{111012000},
 \nonumber \\
 &
 J^{C}_{8}
 & = \;\; & 
 \eps^3
 \left(\frac{m_N^2-s}{\mu^2}\right)
 I^C_{111\left(-1\right)12000}.
\end{alignat}

\subsubsection{Topology D}

A possible choice of master integrals of uniform transcendental weight for topology $D$ is given by:
\begin{alignat}{2}
 \mbox{Sector 50:} \;\;\;\; &
 J^{D}_{1}
 & = \;\; & 
 \eps^2 
 \left(\frac{-t}{\mu^2}\right)
 I^D_{010022000},
 \nonumber \\
 \mbox{Sector 52:} \;\;\;\; &
 J^{D}_{2}
 & = \;\; & 
 \eps \left(1+4\eps\right) 
 \left( \frac{m_N^2}{\mu^2} \right)
 \; {\bf D}^- I^D_{001011000},
 \nonumber \\
 \mbox{Sector 53:} \;\;\;\; &
 J^{D}_{3}
 & = \;\; & 
 -4 \eps^3 
 \left(\frac{m_N^2-s}{\mu^2}\right)
 I^D_{101012000},
 \nonumber \\
 \mbox{Sector 54:} \;\;\;\; &
 J^{D}_{4}
 & = \;\; & 
 \eps^3 
 \left(\frac{r_1}{\mu^2}\right)
 I^D_{011012000},
 \nonumber \\
 \mbox{Sector 55:} \;\;\;\; &
 J^{D}_{5}
 & = \;\; & 
 \eps^3
 \left(\frac{m_N^2-s}{\mu^2}\right) \left(\frac{-t}{\mu^2}\right)
 I^D_{111012000}.
\end{alignat}

\subsubsection{Topology E}

A possible choice of master integrals of uniform transcendental weight for topology $E$ is given by:
\begin{alignat}{2}
 \mbox{Sector 50:} \;\;\;\; &
 J^{E}_{1}
 & = \;\; & 
 \eps^2 
 \left(\frac{m_Z^2-t}{\mu^2}\right)
 \; {\bf D}^- I^E_{010011000},
 \nonumber \\
 &
 J^{E}_{2}
 & = \;\; & 
 \eps^2 
 \; {\bf D}^- I^E_{010\left(-1\right)11000}
 -
 \eps^2 
 \left(\frac{m_Z^2}{\mu^2}\right)
 \; {\bf D}^- I^E_{010011000},
 \nonumber \\
 \mbox{Sector 56:} \;\;\;\; &
 J^{E}_{3}
 & = \;\; & 
 \eps^2 
 \left(\frac{m_Z^2}{\mu^2}\right)
 \; {\bf D}^- I^E_{000111000},
 \nonumber \\
 \mbox{Sector 53:} \;\;\;\; &
 J^{E}_{4}
 & = \;\; & 
 -4 \eps^3 
 \left(\frac{-s}{\mu^2}\right)
 I^E_{101012000},
 \nonumber \\
 \mbox{Sector 58:} \;\;\;\; &
 J^{E}_{5}
 & = \;\; & 
 \eps^2 
 \left(\frac{r_2}{\mu^2}\right) \left(\frac{m_Z^2}{\mu^2}\right)
 \; {\bf D}^- I^F_{010111000}
 -
 \eps^2 
 \left(\frac{r_2}{\mu^2}\right)
 \; {\bf D}^- I^F_{010011000},
 \nonumber \\
 \mbox{Sector 55:} \;\;\;\; &
 J^{E}_{6}
 & = \;\; & 
 \eps^3
 \left(\frac{-s}{\mu^2}\right) \left(\frac{m_Z^2-t}{\mu^2}\right)
 I^E_{111012000},
 \nonumber \\
 &
 J^{E}_{7}
 & = \;\; & 
 \eps^3
 \left(\frac{-s}{\mu^2}\right)
 I^E_{111\left(-1\right)12000}
 -
 \eps^3
 \left(\frac{-s}{\mu^2}\right) \left(\frac{m_Z^2}{\mu^2}\right)
 I^E_{111012000},
 \nonumber \\
 \mbox{Sector 59:} \;\;\;\; &
 J^{E}_{8}
 & = \;\; & 
 \eps^3
 \left(\frac{-t}{\mu^2}\right) \left(\frac{m_Z^2}{\mu^2}\right)
 I^E_{110112000},
 \nonumber \\
 \mbox{Sector 61:} \;\;\;\; &
 J^{E}_{9}
 & = \;\; & 
 \eps^3
 \left(\frac{-s}{\mu^2}\right) \left(\frac{m_Z^2}{\mu^2}\right)
 I^E_{101112000},
 \nonumber \\
 \mbox{Sector 63:} \;\;\;\; &
 J^{E}_{10}
 & = \;\; & 
 \eps^3
 \left(\frac{r_4}{\mu^4}\right) \left(\frac{m_Z^2}{\mu^2}\right)
 I^E_{111112000}
 -
 \eps^3
 \left(\frac{r_4}{\mu^4}\right)
 I^E_{111012000}.
\end{alignat}

\subsubsection{Topology F}

A possible choice of master integrals of uniform transcendental weight for topology $F$ is given by:
\begin{alignat}{2}
 \mbox{Sector 50:} \;\;\;\; &
 J^{F}_{1}
 & = \;\; & 
 \eps^2 
 \left(\frac{-t}{\mu^2}\right)
 I^F_{010022000},
 \nonumber \\
 \mbox{Sector 56:} \;\;\;\; &
 J^{F}_{2}
 & = \;\; & 
 \eps^2 
 \left(\frac{m_Z^2}{\mu^2}\right)
 \; {\bf D}^- I^F_{000111000},
 \nonumber \\
 \mbox{Sector 53:} \;\;\;\; &
 J^{F}_{3}
 & = \;\; & 
 -4 \eps^3 
 \left(\frac{-s}{\mu^2}\right)
 I^F_{101012000},
 \nonumber \\
 \mbox{Sector 58:} \;\;\;\; &
 J^{F}_{4}
 & = \;\; & 
 \eps^2 
 \left(\frac{m_Z^2}{\mu^2}\right)\left(\frac{m_Z^2-t}{\mu^2}\right)
 \; {\bf D}^- I^F_{010111000}
 - \eps^2 
 \left(\frac{m_Z^2}{\mu^2}\right)
 \; {\bf D}^- I^F_{010011000},
 \nonumber \\
 \mbox{Sector 55:} \;\;\;\; &
 J^{F}_{5}
 & = \;\; & 
 \eps^3
 \left(\frac{-s}{\mu^2}\right) \left(\frac{-t}{\mu^2}\right)
 I^F_{111012000},
 \nonumber \\
 \mbox{Sector 61:} \;\;\;\; &
 J^{F}_{6}
 & = \;\; & 
 \eps^3
 \left(\frac{-s}{\mu^2}\right) \left(\frac{m_Z^2}{\mu^2}\right)
 I^F_{101112000},
 \nonumber \\
 \mbox{Sector 63:} \;\;\;\; &
 J^{F}_{7}
 & = \;\; & 
 \eps^3
 \left(\frac{-s}{\mu^2}\right) \left(\frac{m_Z^2-t}{\mu^2}\right) \left(\frac{m_Z^2}{\mu^2}\right)
 I^F_{111112000}
 -
 \eps^3
 \left(\frac{-s}{\mu^2}\right) \left(\frac{m_Z^2}{\mu^2}\right)
 I^F_{111012000}.
\end{alignat}

\subsubsection{Topology G}

A possible choice of master integrals of uniform transcendental weight for topology $G$ is given by:
\begin{alignat}{2}
 \mbox{Sector 50:} \;\;\;\; &
 J^{G}_{1}
 & = \;\; & 
 \eps^2 
 \left(\frac{m_Z^2-t}{\mu^2}\right)
 \; {\bf D}^- I^G_{010011000},
 \nonumber \\
 &
 J^{G}_{2}
 & = \;\; & 
 \eps^2 
 \; {\bf D}^- I^G_{010\left(-1\right)11000},
 \nonumber \\
 \mbox{Sector 53:} \;\;\;\; &
 J^{G}_{3}
 & = \;\; & 
 -4 \eps^3 
 \left(\frac{-s}{\mu^2}\right)
 I^G_{101012000},
 \nonumber \\
 \mbox{Sector 55:} \;\;\;\; &
 J^{G}_{4}
 & = \;\; & 
 \eps^3
 \left(\frac{-s}{\mu^2}\right) \left(\frac{m_Z^2-t}{\mu^2}\right)
 I^G_{111012000},
 \nonumber \\
 &
 J^{G}_{5}
 & = \;\; & 
 \eps^3
 \left(\frac{-s}{\mu^2}\right)
 I^G_{111\left(-1\right)12000}.
\end{alignat}

\subsubsection{Topology H}
 
A possible choice of master integrals of uniform transcendental weight for topology $H$ is given by:
\begin{alignat}{2}
 \mbox{Sector 50:} \;\;\;\; &
 J^{H}_{1}
 & = \;\; & 
 \eps^2 
 \left(\frac{-t}{\mu^2}\right)
 I^H_{010022000},
 \nonumber \\
 \mbox{Sector 53:} \;\;\;\; &
 J^{H}_{2}
 & = \;\; & 
 -4 \eps^3 
 \left(\frac{-s}{\mu^2}\right)
 I^H_{101012000},
 \nonumber \\
 \mbox{Sector 55:} \;\;\;\; &
 J^{H}_{3}
 & = \;\; & 
 \eps^3
 \left(\frac{-s}{\mu^2}\right) \left(\frac{-t}{\mu^2}\right)
 I^H_{111012000}.
\end{alignat}

\section{The differential equations}
\label{sect:differential_equations}

The master integrals $J^X$ satisfy a differential equation in $\eps$-factorised form
\bq
\label{def_differential_equation}
 d J^X & = & M^X J^X,
 \;\;\;\;\;\;
 X \in \left\{ A,B,C,D,E,F,G,H \right\},
\eq
with $M^X$ of the form
\bq
\label{def_M_X}
 M^X & = & \eps \sum\limits_{k=1}^{\NL^X} C_k^X \omega_k^X.
\eq
The differential equation is obtained as follows: 
Standard integration-by-parts reduction programs allow us to obtain the differential equation for the
pre-canonical master integrals $I^X$:
\bq
 d I^X & = & A^X I^X.
\eq
Integration-by-parts reduction allows us also to express any master integral $J^X_i$
of uniform transcendental weight as defined in subsection~\ref{sect:master_integrals_uniform_transcendental_weight} as a linear combination of the pre-canonical master integrals $I^X_j$.
This defines the matrix $U^X$ in eq.~(\ref{def_fibre_transformation}).
The matrix $M^X$ appearing in eq.~(\ref{def_differential_equation}) is then given by
\bq
 M^X & = & U^X A^X \left(U^X\right)^{-1} - U^X d \left(U^X\right)^{-1}.
\eq
The $C_k^X$'s appearing in eq.~(\ref{def_M_X}) are square matrices with constant entries.
The dimension of these matrices is given by the number of master integrals for topology $X$.
The differential one-forms $\omega_k^X$ are of the form
\bq
\label{def_dlog_forms}
 \omega_k^X & = & d \log f_k^X,
\eq
where $f_k^X$ is an algebraic function of $s, t, m_Z^2, m_N^2$ and $\mu^2$.
The $f_k^X$'s are called letters and the set of all $f_k^X$'s is called the alphabet ${\mathcal A}^X$.
The alphabets are
\bq
 {\mathcal A}^A
 & = &
 \left\{ e_1, e_2, e_4, e_6, e_7, e_9, e_{10}, e_{11}, e_{12}, e_{13}, e_{14}, e_{15}, e_{16}, 
  \right. \nonumber \\
  & & \left.
  o_1, o_5, o_6, o_7, o_8, o_9, o_{10}, o_{11}, o_{12}, o_{13}, o_{14}, o_{15}, o_{16} \right\},
 \nonumber \\
 {\mathcal A}^B
 & = &
 \left\{ e_1, e_2, e_4, e_6, e_9, e_{10}, e_{11}, e_{12}, e_{13}, e_{14}, o_5, o_6, o_7, o_8, o_9 \right\},
 \nonumber \\
 {\mathcal A}^C
 & = &
 \left\{ e_1, e_2, e_4, e_6, e_9, e_{10}, e_{11}, e_{12}, e_{13}, e_{14}, o_5, o_6, o_7, o_8, o_9 \right\},
 \nonumber \\
 {\mathcal A}^D
 & = &
 \left\{ e_1, e_2, e_9, e_{10}, e_{11}, e_{12}, o_5, o_6 \right\},
 \nonumber \\
 {\mathcal A}^E
 & = &
 \left\{ e_1, e_2, e_3, e_4, e_5, e_6, e_7, e_8, o_1, o_2, o_3, o_4 \right\},
 \nonumber \\
 {\mathcal A}^F
 & = &
 \left\{ e_1, e_2, e_3, e_4, e_5, e_6 \right\},
 \nonumber \\
 {\mathcal A}^G
 & = &
 \left\{ e_1, e_2, e_3, e_4, e_5, e_6 \right\},
 \nonumber \\
 {\mathcal A}^H
 & = &
 \left\{ e_1, e_2, e_3 \right\}.
\eq
A specific letter may appear in more than one alphabet. 
We divide the letters in rational (or even) letters and non-rational (or odd) letters.
The rational letters are
\begin{align}
 e_1 & = \frac{-s}{\mu^2},
 &
 e_7 & = \frac{4m_Z^2-t}{\mu^2},
 &
 e_{12} & = \frac{\left(m_N^2-s\right)^2+st}{\mu^4},
 \nonumber \\
 e_2 & = \frac{-t}{\mu^2},
 &
 e_8 & = \frac{4m_Z^2\left(m_Z^2+s\right)-st}{\mu^4},
 &
 e_{13} & = \frac{4m_N^2-m_Z^2}{\mu^2},
 \nonumber \\
 e_3 & = \frac{-s-t}{\mu^2},
 &
 e_9 & = \frac{m_N^2}{\mu^2},
 &
 e_{14} & = \frac{\left(m_N^2-s\right)^2+sm_Z^2}{\mu^4},
 \nonumber \\
 e_4 & = \frac{m_Z^2}{\mu^2},
 &
 e_{10} & = \frac{m_N^2-s}{\mu^2},
 &
 e_{15} & = \frac{m_N^2\left(4m_Z^2-t\right)-m_Z^4}{\mu^4},
 \nonumber \\
 e_5 & = \frac{m_Z^2+s}{\mu^2},
 &
 e_{11} & = \frac{4m_N^2-t}{\mu^2},
 &
 e_{16} & = \frac{\left(4m_Z^2-t\right)\left(m_N^2-s\right)^2+ 4 s m_Z^4}{\mu^6}.
 \nonumber \\
 e_6 & = \frac{m_Z^2-t}{\mu^2},
\end{align}
The non-rational letters are 
\bq
 o_1 & = & \frac{2m_Z^2-t-r_2}{2m_Z^2-t+r_2},
 \nonumber \\
 o_2 & = & \frac{2m_Z^2\left(m_Z^2+s\right)-st-r_4}{2m_Z^2\left(m_Z^2+s\right)-st+r_4},
 \nonumber \\
 o_3 & = & \frac{\left(4m_Z^2-t\right)st+2m_Z^4\left(t-s\right)-\left(2m_Z^2-t\right)r_4}{\left(4m_Z^2-t\right)st+2m_Z^4\left(t-s\right)+\left(2m_Z^2-t\right)r_4},
 \nonumber \\
 o_4 & = & \frac{2m_Z^4t+st\left(4m_Z^2-t\right)-r_2r_4}{2m_Z^4t+st\left(4m_Z^2-t\right)+r_2r_4},
 \nonumber \\
 o_5 & = & \frac{2m_N^2-t-r_1}{2m_N^2-t+r_1},
 \nonumber \\
 o_6 & = & \frac{2m_N^2\left(m_N^2-s\right)^2-t\left(m_N^4+s^2\right)-\left(m_N^4-s^2\right)r_1}{2m_N^2\left(m_N^2-s\right)^2-t\left(m_N^4+s^2\right)+\left(m_N^4-s^2\right)r_1},
 \nonumber \\
 o_7 & = & \frac{2m_N^2-m_Z^2-r_3}{2m_N^2-m_Z^2+r_3},
 \nonumber \\
 o_8 & = & \frac{2m_N^2\left(m_Z^2+t\right)-m_Z^2t-r_1r_3}{2m_N^2\left(m_Z^2+t\right)-m_Z^2t+r_1r_3},
 \nonumber \\
 o_9 & = & \frac{2m_N^2\left(m_N^2-s\right)^2-m_Z^2\left(m_N^4+s^2\right)-\left(m_N^4-s^2\right)r_3}{2m_N^2\left(m_N^2-s\right)^2-m_Z^2\left(m_N^4+s^2\right)+\left(m_N^4-s^2\right)r_3},
 \nonumber \\
 o_{10} & = &
 \frac{2m_N^2m_Z^2\left(4m_N^2-m_Z^2\right)+\left[2m_N^4-m_Z^2\left(4m_N^2-m_Z^2\right)\right]t-\left(2m_N^2-m_Z^2\right) r_1 r_3}{2m_N^2m_Z^2\left(4m_N^2-m_Z^2\right)+\left[2m_N^4-m_Z^2\left(4m_N^2-m_Z^2\right)\right]t+\left(2m_N^2-m_Z^2\right) r_1 r_3},
 \nonumber \\
 o_{11} & = &
 \frac{-t\left[2m_Z^4+\left(2m_N^2-t\right)\left(4m_Z^2-t\right)\right]-\left(2m_Z^2-t\right)r_1r_2}{-t\left[2m_Z^4+\left(2m_N^2-t\right)\left(4m_Z^2-t\right)\right]+\left(2m_Z^2-t\right)r_1r_2},
 \nonumber \\
 o_{12} & = &
 \frac{\left(m_N^2-s\right)^2\left(2m_Z^2-t\right) + 2 s m_Z^4 - \left(m_N^2-s\right) r_5}{\left(m_N^2-s\right)^2\left(2m_Z^2-t\right) + 2 s m_Z^4 + \left(m_N^2-s\right) r_5},
 \nonumber \\
 o_{13} & = &
 \frac{\left(m_N^2-s\right)^2\left(2m_z^4-4m_Z^2t+t^2\right)-2m_Z^4st-\left(m_N^2-s\right)\left(2m_Z^2-t\right)r_5}{\left(m_N^2-s\right)^2\left(2m_z^4-4m_Z^2t+t^2\right)-2m_Z^4st+\left(m_N^2-s\right)\left(2m_Z^2-t\right)r_5},
 \nonumber \\
 o_{14} & = &
 \frac{-t Q_8-\left(m_N^2+s\right)\left(2m_Z^2-t\right)r_1 r_5}{-t Q_8+\left(m_N^2+s\right)\left(2m_Z^2-t\right)r_1 r_5},
 \nonumber \\
 o_{15} & = &
 \frac{-t\left(m_N^2+s\right)m_Z^2-r_3r_5}{-t\left(m_N^2+s\right)m_Z^2+r_3r_5},
 \nonumber \\
 o_{16} & = &
 \frac{-t\left[2m_Z^4-\left(m_N^2-s\right)\left(4m_Z^2-t\right)\right]-r_2r_5}{-t\left[2m_Z^4-\left(m_N^2-s\right)\left(4m_Z^2-t\right)\right]+r_2r_5},
\eq
with
\bq
 Q_8
 & = &
 \left(4m_Z^2-t\right) \left[ 2m_N^2\left(m_N^2-s\right)^2-t\left(m_N^4+s^2\right)\right] 
 +2m_Z^4\left(\left(m_N^2+s\right)^2+s\left(4m_N^2-t\right)\right).
\eq
As an example let us write down the differential equation for topology $H$:
\bq
 d J^H & = & M^H J^H,
 \\
 M^H & = & 
 \eps \left( \begin{array}{rrr}
        0 & 0 & 0 \\
        0 & -2 & 0 \\
        0 & - \frac{3}{4} & -1 \\
      \end{array} \right) d \log e_1
 +
 \eps \left( \begin{array}{rrr}
        -2 & 0 & 0 \\
        0 & 0 & 0 \\
        -\frac{3}{2} & 0 & -2 \\
      \end{array} \right) d \log e_2
 +
 \eps \left( \begin{array}{rrr}
        0 & 0 & 0 \\
        0 & 0 & 0 \\
        \frac{3}{2} & \frac{3}{4} & 1 \\
      \end{array} \right) d \log e_3.
 \nonumber
\eq
The differential equations for the other topologies are of a similar form.
The matrices $M^X$ for all topologies are given in an electronic file
attached to the arxiv version of this article.

\section{Rationalisation of the square roots}
\label{sect:rationalisation}

The topologies $A$-$E$ contain square roots.
For each topology all occurring square roots can be rationalised simultaneously.
This implies that all integrals can be expressed in terms of multiple polylogarithms.
For the rationalisation of the square roots we use the algorithms of \cite{Besier:2018jen,Besier:2019kco}.

\subsection{The square root $r_1$}
\label{subsection:r1}

Topology $D$ involves only the square root $r_1$.
The standard rationalisation of the square root $r_1$ is given by
\bq
 t = - \frac{\left(1-y\right)^2}{y} m_N^2,
 & &
 r_1 = \frac{1-y^2}{y} m_N^2,
\eq
with the inverse transformation given by
\bq
 y \; = \; \frac{1}{2} \left( 2-\frac{t}{m_N^2} - \frac{r_1}{m_N^2} \right).
\eq
It is convenient to use instead of $y$ the variable $\bar{y}=1-y$.
This has the advantage that $t=0$ corresponds to $\bar{y}=0$.
In integrating the differential equation we will choose $t=0$ as boundary.
In terms of $\bar{y}$ we have
\bq
 t = - \frac{\bar{y}^2}{1-\bar{y}} m_N^2,
 & &
 r_1 = \frac{\bar{y}\left(2-\bar{y}\right)}{1-\bar{y}} m_N^2.
\eq
The inverse transformation is given by
\bq
 \bar{y} \; = \; \frac{t+r_1}{2m_N^2}.
\eq

\subsection{The square roots $r_1$ and $r_3$}
\label{subsection:r1r3}

Topologies $B$ and $C$ involve the square roots $r_1$ and $r_3$. 
The square root $r_1$ is rationalised as above, the square root $r_3$ is 
rationalised by
\bq
\label{eq_mZ2_to_z}
 m_Z^2 = \frac{\left(1+z\right)^2}{z} m_N^2,
 & &
 r_3 = \frac{1-z^2}{z} m_N^2.
\eq
The inverse transformation is given by
\bq
 z \; = \; -\frac{1}{2} \left( 2-\frac{m_Z^2}{m_N^2} + \frac{r_3}{m_N^2} \right).
\eq

\subsection{The square roots $r_2$ and $r_4$}
\label{subsection:r2r4}

Topology $E$ involves the square roots $r_2$ and $r_4$.
The square root $r_2$ is rationalised by
\bq
 t = - \frac{\tilde{y}^2}{1-\tilde{y}} m_Z^2,
 & &
 r_2 = \frac{\tilde{y}\left(2-\tilde{y}\right)}{1-\tilde{y}} m_Z^2.
\eq
The inverse transformation is given by
\bq
 \tilde{y} \; = \; \frac{t+r_2}{2 m_Z^2}.
\eq
The square root $r_4$ is rationalised by
\bq
 m_Z^2 = \frac{\left(1-2\tilde{z}\right)}{4\tilde{z}^2} \frac{\left(2-\tilde{y}\right)^2}{\left(1-\tilde{y}\right)} s,
 & &
 r_4 = 
 \frac{\left(1-\tilde{z}\right)\left(1-2\tilde{z}\right)}{4\tilde{z}^3}
 \frac{\tilde{y}\left(2-\tilde{y}\right)^3}{\left(1-\tilde{y}\right)^2} s^2.
\eq
The inverse transformation is given by
\bq
 \tilde{z} \; = \; 
 -\frac{\left(4m_Z^2-t\right)}{4 m_Z^2} \left( \frac{s}{m_Z^2} - \frac{r_4}{m_Z^2 r_2} \right).
\eq

\subsection{The square roots $r_1$, $r_2$, $r_3$ and $r_5$}
\label{subsection:r1r2r3r5}

Topology $A$ involves the square roots $r_1$, $r_2$ $r_3$ and $r_5$.
The square roots $r_1$ and $r_2$ are rationalised by
\bq
\label{def_rationalise_r1r2}
 t
 & = &
 - \frac{\left(\hat{y}^2m_Z^2-m_N^2\right)^2}{\hat{y}\left(1+\hat{y}\right)\left(m_N^2+\hat{y}m_Z^2\right)},
 \nonumber \\
 r_1
 & = &
 \frac{\left(\hat{y}^2m_Z^2-m_N^2\right)\left(\left(1+2\hat{y}\right)m_N^2+\hat{y}^2m_Z^2\right)}{\hat{y}\left(1+\hat{y}\right)\left(m_N^2+\hat{y}m_Z^2\right)},
 \nonumber \\
 r_2
 & = &
 \frac{\left(\hat{y}^2m_Z^2-m_N^2\right)\left(m_N^2+\hat{y}\left(2+\hat{y}\right)m_Z^2\right)}{\hat{y}\left(1+\hat{y}\right)\left(m_N^2+\hat{y}m_Z^2\right)}.
\eq
The inverse transformation is given by
\bq
 \hat{y}
 \; = \;
 -\frac{t^2 - t\left(r_1+r_2\right)+r_1r_2}{4m_Z^2 t}.
\eq
The root $r_3$ is rationalised as in section~\ref{subsection:r1r3}:
\bq
 m_Z^2 = \frac{\left(1+z\right)^2}{z} m_N^2,
 & &
 r_3 = \frac{1-z^2}{z} m_N^2.
\eq
The inverse transformation is given by
\bq
 z \; = \; -\frac{1}{2} \left( 2-\frac{m_Z^2}{m_N^2} + \frac{r_3}{m_N^2} \right).
\eq
Finally, the root $r_5$ is rationalised by
\bq
 s & = &
 \frac{\hat{x}}{\left(1+\hat{x}\right)}
 \left( m_N^2 + \hat{x} \frac{m_Z^4}{4m_Z^2-t} \right),
 \nonumber \\
 r_5 & = &
 \frac{-t}{\left(1+\hat{x}\right)r_2}
 \left[ m_N^2 \left(4m_Z^2-t\right) + \hat{x} \left(2+\hat{x}\right) m_Z^4 \right].
\eq
The occurrence of $r_2$ on the right-hand side is unproblematic, as $r_2$ is rationalised by eq.~(\ref{def_rationalise_r1r2}).
The inverse transformation is given by
\bq
 \hat{x} 
 & = &
- \frac{1}{2m_Z^4} \left[ \left(m_N^2-s\right)\left(4m_Z^2-t\right) + \frac{r_2 r_5}{t} \right].
\eq

\section{Boundary values}
\label{sect:boundary}

In order to solve the differential equation, we need boundary values.
As boundary point we choose
\bq
 t \; = \; 0,
 \;\;\;\;\;\;
 m_N^2 \; = \; 0,
 \;\;\;\;\;\;
 m_Z^2 \; = \; \infty.
\eq
There are four master integrals with a trivial kinematic dependence. 
These integrals are easily calculated with the help of the
Feynman parametrisation.
We find
\bq
 J^H_1
 & = &
 e^{2\eps \Eulerconstant}
 \left(\frac{-t}{\mu^2}\right)^{-2\eps}
 \frac{\Gamma\left(1+2\eps\right)\Gamma\left(1-\eps\right)^3}{\Gamma\left(1-3\eps\right)},
 \nonumber \\
 J^H_2
 & = &
 e^{2\eps \Eulerconstant}
 \left(\frac{-s}{\mu^2}\right)^{-2\eps}
 \frac{\Gamma\left(1+2\eps\right)\Gamma\left(1+\eps\right)\Gamma\left(1-2\eps\right)\Gamma\left(1-\eps\right)^2}{\Gamma\left(1-3\eps\right)},
 \nonumber \\
 J^F_2
 & = &
 e^{2\eps \Eulerconstant}
 \left(\frac{m_Z^2}{\mu^2}\right)^{-2\eps}
 \Gamma\left(1+2\eps\right)\Gamma\left(1+\eps\right)\Gamma\left(1-\eps\right),
 \nonumber \\
 J^D_2
 & = &
 \frac{3}{2}
 e^{2\eps \Eulerconstant}
 \left(\frac{m_N^2}{\mu^2}\right)^{-2\eps}
 \frac{\Gamma\left(1+2\eps\right)\Gamma\left(1+\eps\right)\Gamma\left(1-\eps\right)^2\Gamma\left(1-4\eps\right)}{\Gamma\left(1-2\eps\right)\Gamma\left(1-3\eps\right)}.
\eq
The boundary values of the master integrals of intermediate complexity we compute with the help of the Mellin-Barnes representation.
We illustrate this for the example of the master integral $J^F_6$.
For this integral we obtain the Mellin-Barnes representation
\bq
 J^F_6
 & = &
 - e^{2\eps \Eulerconstant} \eps^2
 \frac{\Gamma\left(1+\eps\right)\Gamma\left(1-\eps\right)^2}{\Gamma\left(1-2\eps\right)}
 \left(\frac{m_Z^2}{\mu^2}\right)^{-2\eps}
 \nonumber \\
 & &
 \frac{1}{2\pi i} \int d\sigma
 \left(\frac{-s}{m_Z^2}\right)^{\sigma+1}
 \frac{\Gamma\left(-\sigma\right)\Gamma\left(-\sigma-1-2\eps\right)\Gamma\left(\sigma+1\right)^2\Gamma\left(\sigma+2+2\eps\right)}{\Gamma\left(\sigma+2-\eps\right)}.
\eq
The integration contour runs from $-i\infty$ to $+i\infty$ 
and separates the poles of $\Gamma(-\sigma)$ and $\Gamma(-\sigma-1-2\eps)$ from the poles of $\Gamma(\sigma+1)$ and $\Gamma(\sigma+2+2\eps)$.
For $|s| < m_Z^2$ we may close the integration contour to the right and sum up the residues 
of $\Gamma(-\sigma)$ and $\Gamma(-\sigma-1-2\eps)$.
For the boundary value we are only interested in the leading term in an expansion in $1/m_Z^2$.
The leading term is given by the first residue of $\Gamma(-\sigma-1-2\eps)$ located at
\bq
 \sigma & =& -1-2\eps.
\eq
We therefore obtain
\bq
 J^F_6
 & = &
 - \frac{1}{4} e^{2\eps \Eulerconstant}
 \frac{\Gamma\left(1+2\eps\right)\Gamma\left(1+\eps\right)\Gamma\left(1-\eps\right)^2\Gamma\left(1-2\eps\right)}{\Gamma\left(1-3\eps\right)}
 \left(\frac{-s}{\mu^2}\right)^{-2\eps}
 +{\mathcal O}\left(\frac{-s}{m_Z^2}\right).
\eq
The boundary values of the more complicated integrals we obtain from regularity conditions.
For example, the boundary values for $J^H_3$ are determined from the condition that this integral is regular at $e_3=0$, this corresponds to the 
condition that there is no singularity whenever the Mandelstam variable $u$ vanishes.
This follows from physics:
As the master integral is a planar integral, there is no singularity in the crossed $u$-channel.

\section{Results}
\label{sect:results}

We set $\mu^2=s$.
The values of the Feynman integrals at another scale $\mu_1^2$ are easily obtained through
\bq
 J\left(\mu_1^2\right)
 \; = \;
 e^{2 \eps L}
 J\left(\mu^2\right)
 & \mbox{with} &
 L \; = \; \ln\left(\frac{\mu_1^2}{\mu^2}\right).
\eq

\subsection{Integrating the differential equation}

As already mentioned in section~\ref{sect:boundary}, we use
\bq
 t \; = \; 0,
 \;\;\;\;\;\;
 m_N^2 \; = \; 0,
 \;\;\;\;\;\;
 m_Z^2 \; = \; \infty
\eq
as boundary point.
After setting $\mu^2=s$ the Feynman integrals depend only on dimensionless kinematic variables, which
we may take as
\begin{align}
 x_t & = \frac{-t}{s},
 &
 x_{m_N^2} & = \frac{m_N^2}{s},
 &
 x^{-1}_{m_Z^2} & = \frac{s}{m_Z^2}.
\end{align}
Our chosen boundary point corresponds to
\bq
 \left( x_t, x_{m_N^2}, x^{-1}_{m_Z^2} \right)
 & = &
 \left( 0, 0, 0 \right).
\eq
The rationalisation of square roots will introduce a change of variables.
By a suitable definition of the new variables we may ensure that the boundary point in the new variables is still
$(0, 0, 0)$.
Let us now assume that our integration variables are $(x_1,x_2,x_3)$.
We then fix an integration path $\gamma$: We first integrate along $x_1$ at $x_2=x_3=0$, followed
by an integration along $x_2$ at $x_1=\mathrm{const}$ and $x_3=0$ and finally an integration along $x_3$
at $x_1=\mathrm{const}$ and $x_2=\mathrm{const}$.

\subsubsection{Integration for topologies $B$-$H$}

In table~\ref{table_integration_order} we show for topologies $B$-$H$ the
dimensionless kinematic variables and the integration order.
\begin{table}
\begin{center}
\begin{tabular}{|c|c|}
\hline
 Topologies & Variables and integration order \\
\hline
 $B$, $C$ & $\left( x_{m_N^2}, z, \bar{y} \right)$ \\
 $D$ & $\left( x_{m_N^2}, \bar{y} \right)$ \\
 $E$ & $\left(\tilde{z}, \tilde{y}\right)$ \\
 $F$, $G$ & $\left( x^{-1}_{m_Z^2}, x_t \right)$ \\
 $H$ & $\left( x_t \right)$ \\
\hline
\end{tabular}
\end{center}
\caption{
The dimensionless kinematic variables and the integration order for topologies $B$-$H$. 
}
\label{table_integration_order}
\end{table}
For topologies $B$ and $C$ we use the variable $\bar{y}$ from section~\ref{subsection:r1}
and the variable $z$ from section~\ref{subsection:r1r3}.
The variable $\bar{y}$ from section~\ref{subsection:r1}
is also used for topology $D$.
For topology $E$ we use the variables $\tilde{y}$ and $\tilde{z}$ from section~\ref{subsection:r2r4}.
The rationalisation of the square roots turns the arguments of the dlog-forms given in eq.~(\ref{def_dlog_forms})
into rational functions.
This implies that all iterated integrals from the integration of the differential equation can be expressed in terms
of multiple polylogarithms.
Multiple polylogarithms are defined as follows:
One first defines $G(0,...,0;y)$ with $k$ zeros to be
\bq
 G(0,\dots,0;y)
 & = &
 \frac{1}{k!} \left( \ln y \right)^k.
\eq
This includes the trivial case $G(;y)=1$.
Multiple polylogarithms are then defined recursively by
\bq
\label{Grecursive}
 G(z_1,z_2,\dots,z_k;y) & = & \int\limits_0^y \frac{dy_1}{y_1-z_1} G(z_2,\dots,z_k;y_1).
\eq
A multiple polylogarithm $G(z_1,\dots,z_k;y)$ is said to have a trailing zero if $z_k=0$.
Using the shuffle product, we may isolate trailing zeros in powers of
\bq
 G(0;y)
 & = &
 \ln y.
\eq
An example is given by
\bq
 G(z,0;y)
 & = &
 \ln y \cdot G(z;y) - G(0,z;y).
\eq
For multiple polylogarithms without a trailing zero we have the scaling identity
\bq
 G(z_1,\dots,z_k;y)
 & = &
 G(\frac{z_1}{y},\dots,\frac{z_k}{y};1)
\eq
A multiple polylogarithm $G(z_1,\dots,z_k;1)$ has a convergent power series expansion if
$z_1 \neq 1$ and
\bq
 \left| z_j \right| \; \ge \; 1 & & \mbox{for all} \; j \; \in \; \{1,\dots,k\}.
\eq
The kinematic variables and the integration order for topologies $B$-$H$ have been chosen such that 
after isolating all trailing zeros in powers of logarithms, the remaining multiple polylogarithms have
convergent power series expansions in the kinematic region of interest.
This ensures a highly efficient numerical evaluation.
Furthermore, since we have for small values of $(-t)$ 
\bq
 -t \; \sim \; \bar{y}^2 \; \sim \; \tilde{y}^2
\eq
and since we have for large values of $m_Z^2$
\bq
 m_Z^2 \; \sim \; z^{-1} \; \sim \; \tilde{z}^{-2}
\eq
this procedure also makes the large logarithms
\bq
 \ln\left(\frac{-t}{s}\right)
 & \mbox{and} &
 \ln\left(\frac{s}{m_Z^2}\right)
\eq
manifest.
As an example one finds
\bq
\lefteqn{
 J^F_7
 =
 -\frac{5}{4}
 + \left( 2 L_t -\frac{1}{2} i \pi \right) \eps
 + \left( -L_t^2 + 2 i \pi L_t + \frac{15}{4} \zeta_2 \right) \eps^2
 + \left( -2 i \pi L_t^2 - 12 \zeta_2 L_t + \frac{89}{6} \zeta_3 
 \right.
} & & \nonumber \\
 & &
 \left.
          - \frac{1}{12} i \pi^3 \right) \eps^3
 + \left[ \frac{1}{3} L_t^4 + \frac{4}{3} i \pi L_t^3 + 11 \zeta_2 L_t^2 - \left( \frac{70}{3} \zeta_3 + \frac{1}{3} i \pi^3 \right) L_t
          + \frac{493}{16} \zeta_4 + \frac{19}{3} i \pi \zeta_3 \right] \eps^4 
 \nonumber \\
 & &
 + {\mathcal O}\left(x_{m_Z^2}^{-1}\right)
 + {\mathcal O}\left(x_t\right)
 + {\mathcal O}\left(\eps^5\right),
\eq
with $L_t=\ln(-t/s)$.

\subsubsection{Integration for topology $A$}

For topologies $B$-$H$ we were in the lucky situation that with our choice of variables 
all multiple polylogarithms had convergent power series expansions in the kinematic region of interest.
This is no longer straightforward for topology $A$.

The variables $\hat{x}, \hat{y}, z$ from section~\ref{subsection:r1r2r3r5}  rationalise all square roots
for topology $A$. 
This allows us to conclude that all integrals from topology $A$ can be expressed in terms of multiple polylogarithms
and that all master integrals from topology $A$ are of uniform weight. 
The latter statement uses the fact that the boundary constants are of uniform weight as well \cite{Frellesvig:2023iwr}.
As the master integrals $J^A_1$-$J^A_{13}$ appear also in other topologies, the new information 
is the statement that the master integrals $J^A_{14}$ and $J^A_{15}$ can be  expressed in term of multiple polylogarithms 
and are of uniform weight.

The polynomials appearing in the dlog-forms after rationalisation suggest the integration order
\bq
 \left( z, \hat{y}, \hat{x} \right).
\eq
This integration order has the property that for each integration we encounter at most quadratic polynomials in the integration
variable.
However, the resulting expression in terms of multiple polylogarithms involves multiple polylogarithms which do not
have convergent power series expansions in the kinematic region of interest.
This is not a fundamental problem, as we may transform these multiple polylogarithms 
such that they do have a convergent power series expansion.
However, these transformations are rather slow.

Choosing other integration orders will result in polynomials of higher degree in the integration variables.
While we may determine the roots of these polynomials numerically, there is a second drawback: Factorising these polynomials into linear
factors will significantly increase the number of terms.
To give an example: Consider an iterated integral of depth $w$, where each dlog-form has as argument a polynomial of degree $N$ in the integration
variable.
A single iterated integral of this type will lead to $N^w$ terms.
This growth prevents an efficient numerical evaluation.

An efficient numerical evaluation routine for the kinematic region of interest 
is achieved as follows:
We introduce the variable
\bq
 v & = & \sqrt{x_t}
\eq
and integrate the differential equation with the integration order
\bq
 \left(x_{m_N^2}, z, v \right).
\eq
The first two integrations are done at $t=0$ and give multiple polylogarithms.
From the definition of the square roots in eq.~(\ref{def_square_roots}) we see that at $t=0$
only the square root $r_3$ is non-zero.
The square root $r_3$ is rationalised by the change of variables from $m_Z^2$ to $z$ given in eq.~(\ref{eq_mZ2_to_z}).
There is one subtlety: The $(t \rightarrow 0)$-limit of the letters $o_{11}$, $o_{14}$ and $o_{16}$ is not a  
rational function of $x_{m_N^2}$ and $z$.
However, in the differential equation these letters always multiply the integrals $J^A_6$, $J^A_8$ and $J^A_{14}$.
These integrals vanish at $t=0$ and as a consequence the letters $o_{11}$, $o_{14}$ and $o_{16}$ do not appear
in the result for the master integrals at $t=0$.

For the last integration in $v$ we have the integration kernels
\bq
 \left\{ e_2, e_6, e_7, e_{11}, e_{12}, e_{15}, e_{16}, o_1, o_5, o_6, o_8, o_{10}, o_{11}, o_{12}, o_{13}, o_{14}, o_{15}, o_{16} \right\}.
\eq
$d\ln(e_2)$ has a simple pole at $v=0$, all others have Taylor expansions in $v$.
The Taylor expansions are convergent for 
\bq
 \left|t\right| < 
 \min\left\{ 4 m_N^2, m_Z^2, \left|\frac{m_Z^2\left(4m_N^2-m_Z^2\right)}{m_N^2}\right|,
 \left| \frac{\left(m_N^2-s\right)^2}{s} \right|,
 \left| \frac{4m_Z^2\left[\left(m_N^2-s\right)^2+sm_Z^2\right]}{\left(m_N^2-s\right)^2} \right| \right\}.
\eq
This includes the kinematic region of the P2 experiment.
For large $m_Z^2$, the most stringent condition is
\bq
 \left|t\right| & < & \left| \frac{\left(m_N^2-s\right)^2}{s} \right|.
\eq
This is nothing else than the condition on the physical region \cite{Byckling:1971vca}.

\subsection{Numerical results}
\label{sect:numerical_results}

For the numerical evaluation we have written for each topology a {\tt C++}-program, which uses the {\tt GiNaC}-library \cite{Bauer:2000cp}.
This allows numerical evaluations with arbitrary precision.
The algorithms for the numerical evaluation of multiple polylogarithms are based on \cite{Vollinga:2004sn},
topology $A$ uses in addition the class \verb|user_defined_kernel| from ref.~\cite{Walden:2020odh} for the last integration over the variable $v$.

As a typical kinematical point we use
\begin{align}
\label{def_kinematic_point}
 s & = 1.18 \; \mathrm{GeV}^2,
 &
 t & = - 4.5 \cdot 10^{-3} \; \mathrm{GeV}^2,
 &
 m_Z^2 & = 8.32 \cdot 10^{3} \; \mathrm{GeV}^2,
 & 
 m_N^2 & = 0.867 \; \mathrm{GeV}^2.
\end{align}
This corresponds to the kinematics of the P2 experiment with an electron beam of $E_{\mathrm{beam}} = 155 \; \mathrm{MeV}$ and
momentum transfer of $Q^2=-t= 4.5 \cdot 10^{-3} \; \mathrm{GeV}^2$.
The Mandelstam variable $s$ is then given by
\bq
 s & = &
 \left( E_{\mathrm{beam}} + m_N \right)^2.
\eq
We use 
\bq
 m_N \; = \; 0.931 \; \mathrm{GeV},
 & &
 m_Z \; = \; 91.2 \; \mathrm{GeV}.
\eq
The values of the master integrals for the first five terms of the $\eps$-expansion
at the kinematic point specified by eq.~(\ref{def_kinematic_point})
are given to $8$ digits in tables~\ref{table_numerical_results_topo_A} to~\ref{table_numerical_results_topo_H}.
In addition, we compared our results to the results of the program \verb|AMFlow| \cite{Liu:2022chg,Liu:2017jxz,Liu:2022mfb}
to $50$ digits and found perfect agreement.
Our numerical evaluation routines are significantly faster than \verb|AMFlow|.
For example, our program takes about eight seconds to evaluate all master integrals of topology $A$ to $50$ digits 
up to and including the $\eps^4$-term.
The corresponding evaluation with \verb|AMFlow| takes about $21$ minutes.
All calculations were done on a single core of a standard desktop computer.
\begin{table}[!htbp]
\begin{center}
{\small
\begin{tabular}{|l|lllll|}
 \hline 
 & $\eps^0$ & $\eps^1$ & $\eps^2$ & $\eps^3$ & $\eps^4$ \\
 \hline 
$J^{A}_{ 1}$ & $        1$ & $-17.721808$ & $ 161.96604$ & $-1018.2849$ & $ 4958.5733$ \\ 
$J^{A}_{ 2}$ & $       -1$ & $ 17.721806$ & $-161.96601$ & $ 1018.2846$ & $-4958.5714$ \\ 
$J^{A}_{ 3}$ & $      1.5$ & $ 0.92469222$ & $ 12.622024$ & $ 20.886482$ & $ 133.48343$ \\ 
$J^{A}_{ 4}$ & $        1$ & $-17.721806$ & $ 161.96601$ & $-1018.2846$ & $ 4958.5715$ \\ 
$J^{A}_{ 5}$ & $      0.5$ & $ 2.3459023$ & $-20.695763$ & $-152.3306$ & $-545.47467$ \\ 
 &  & $+  6.2831853 i$ & $+  33.35282 i$ & $+  57.516544 i$ & $-73.412472 i$ \\ 
$J^{A}_{ 6}$ & $        0$ & $        0$ & $-0.00015280111$ & $ 0.00094650525$ & $-0.0070979484$ \\ 
$J^{A}_{ 7}$ & $        0$ & $-18.337851$ & $ 153.5486$ & $-1032.1747$ & $ 4869.4159$ \\ 
$J^{A}_{ 8}$ & $        0$ & $ 0.002941742$ & $-0.052132982$ & $ 0.47646222$ & $-2.9955308$ \\ 
$J^{A}_{ 9}$ & $        0$ & $ 18.337851$ & $-153.54859$ & $ 1032.1747$ & $-4869.4155$ \\ 
$J^{A}_{ 10}$ & $   -0.125$ & $-0.58647531$ & $  5.17355$ & $ 38.08511$ & $ 136.35187$ \\ 
 &  & $-1.5707963 i$ & $-8.3382329 i$ & $-14.379368 i$ & $+  18.352165 i$ \\ 
$J^{A}_{ 11}$ & $        0$ & $        0$ & $ 0.00039275926$ & $-0.0024636613$ & $ 0.016696228$ \\ 
 &  &  & $+  3.1349531e-05 i$ & $+  0.00026046032 i$ & $+  0.0010683564 i$ \\ 
$J^{A}_{ 12}$ & $        0$ & $ 2.7043262e-07$ & $-3.9812567e-06$ & $ 3.0234529e-05$ & $-0.00015754108$ \\ 
$J^{A}_{ 13}$ & $   -0.125$ & $-0.58647558$ & $ 5.1735479$ & $ 38.085115$ & $ 136.35197$ \\ 
 &  & $-1.5707963 i$ & $-8.3382363 i$ & $-14.379397 i$ & $+  18.35205 i$ \\ 
$J^{A}_{ 14}$ & $        0$ & $        0$ & $-0.00013778848$ & $ 0.00069534307$ & $-0.0049161254$ \\ 
$J^{A}_{ 15}$ & $        0$ & $        0$ & $ 0.00016537254$ & $-0.00085429938$ & $ 0.0052303379$ \\ 
 &  &  & $+  1.4597012e-05 i$ & $+  0.00012127593 i$ & $+  0.00049744992 i$ \\ 
 \hline 
\end{tabular}
}
\end{center}
\caption{
Numerical results for the first five terms of the $\eps$-expansion of the master integrals $J^A_{1}$-$J^A_{15}$ 
for the kinematic point specified by eq.~(\ref{def_kinematic_point}).
}
\label{table_numerical_results_topo_A}
\end{table}
\begin{table}[!htbp]
\begin{center}
{\small
\begin{tabular}{|l|lllll|}
 \hline 
 & $\eps^0$ & $\eps^1$ & $\eps^2$ & $\eps^3$ & $\eps^4$ \\
 \hline 
$J^{B}_{ 1}$ & $        1$ & $ 11.138385$ & $ 60.386872$ & $ 199.16752$ & $ 381.04712$ \\ 
$J^{B}_{ 2}$ & $      1.5$ & $ 0.92469222$ & $ 12.622024$ & $ 20.886482$ & $ 133.48343$ \\ 
$J^{B}_{ 3}$ & $        1$ & $-17.721806$ & $ 161.96601$ & $-1018.2846$ & $ 4958.5715$ \\ 
$J^{B}_{ 4}$ & $      0.5$ & $ 2.3459023$ & $-20.695763$ & $-152.3306$ & $-545.47467$ \\ 
 &  & $+  6.2831853 i$ & $+  33.35282 i$ & $+  57.516544 i$ & $-73.412472 i$ \\ 
$J^{B}_{ 5}$ & $        0$ & $        0$ & $-9.3465893$ & $-122.24172$ & $-894.31476$ \\ 
$J^{B}_{ 6}$ & $        1$ & $ 39.998578$ & $-41.192268$ & $  1416.62$ & $-4196.4777$ \\ 
$J^{B}_{ 7}$ & $        0$ & $ 18.337851$ & $-153.54859$ & $ 1032.1747$ & $-4869.4155$ \\ 
$J^{B}_{ 8}$ & $   -1.125$ & $-10.113215$ & $-34.840045$ & $-15.424624$ & $ 354.88999$ \\ 
 &  & $-4.712389 i$ & $-52.488401 i$ & $-282.51753 i$ & $-923.44496 i$ \\ 
$J^{B}_{ 9}$ & $   -0.125$ & $-0.58647558$ & $ 5.1735479$ & $ 38.085115$ & $ 136.35197$ \\ 
 &  & $-1.5707963 i$ & $-8.3382363 i$ & $-14.379397 i$ & $+  18.35205 i$ \\ 
$J^{B}_{ 10}$ & $        0$ & $        0$ & $-9.3464365$ & $-122.24267$ & $-894.30774$ \\ 
$J^{B}_{ 11}$ & $   -0.625$ & $-6.1556684$ & $-28.400251$ & $-48.368222$ & $ 100.24132$ \\ 
 &  & $-1.5707963 i$ & $-26.654035 i$ & $-173.96566 i$ & $-633.98226 i$ \\ 
 \hline 
\end{tabular}
}
\end{center}
\caption{
Numerical results for the first five terms of the $\eps$-expansion of the master integrals $J^B_{1}$-$J^B_{11}$ 
for the kinematic point specified by eq.~(\ref{def_kinematic_point}).
}
\label{table_numerical_results_topo_B}
\end{table}
\begin{table}[!htbp]
\begin{center}
{\small
\begin{tabular}{|l|lllll|}
 \hline 
 & $\eps^0$ & $\eps^1$ & $\eps^2$ & $\eps^3$ & $\eps^4$ \\
 \hline 
$J^{C}_{ 1}$ & $        1$ & $-17.721808$ & $ 161.96604$ & $-1018.2849$ & $ 4958.5733$ \\ 
$J^{C}_{ 2}$ & $       -1$ & $ 17.721806$ & $-161.96601$ & $ 1018.2846$ & $-4958.5714$ \\ 
$J^{C}_{ 3}$ & $      1.5$ & $ 0.92469222$ & $ 12.622024$ & $ 20.886482$ & $ 133.48343$ \\ 
$J^{C}_{ 4}$ & $      0.5$ & $ 2.3459023$ & $-20.695763$ & $-152.3306$ & $-545.47467$ \\ 
 &  & $+  6.2831853 i$ & $+  33.35282 i$ & $+  57.516544 i$ & $-73.412472 i$ \\ 
$J^{C}_{ 5}$ & $        0$ & $        0$ & $-0.00015280111$ & $ 0.00094650525$ & $-0.0070979484$ \\ 
$J^{C}_{ 6}$ & $        0$ & $-18.337851$ & $ 153.5486$ & $-1032.1747$ & $ 4869.4159$ \\ 
$J^{C}_{ 7}$ & $   -0.125$ & $-0.58647531$ & $  5.17355$ & $ 38.08511$ & $ 136.35187$ \\ 
 &  & $-1.5707963 i$ & $-8.3382329 i$ & $-14.379368 i$ & $+  18.352165 i$ \\ 
$J^{C}_{ 8}$ & $        0$ & $        0$ & $ 0.00039275926$ & $-0.0024636613$ & $ 0.016696228$ \\ 
 &  &  & $+  3.1349531e-05 i$ & $+  0.00026046032 i$ & $+  0.0010683564 i$ \\ 
 \hline 
\end{tabular}
}
\end{center}
\caption{
Numerical results for the first five terms of the $\eps$-expansion of the master integrals $J^C_{1}$-$J^C_{8}$ 
for the kinematic point specified by eq.~(\ref{def_kinematic_point}).
}
\label{table_numerical_results_topo_C}
\end{table}
\begin{table}[!htbp]
\begin{center}
{\small
\begin{tabular}{|l|lllll|}
 \hline 
 & $\eps^0$ & $\eps^1$ & $\eps^2$ & $\eps^3$ & $\eps^4$ \\
 \hline 
$J^{D}_{ 1}$ & $        1$ & $ 11.138385$ & $ 60.386872$ & $ 199.16752$ & $ 381.04712$ \\ 
$J^{D}_{ 2}$ & $      1.5$ & $ 0.92469222$ & $ 12.622024$ & $ 20.886482$ & $ 133.48343$ \\ 
$J^{D}_{ 3}$ & $      0.5$ & $ 2.3459023$ & $-20.695763$ & $-152.3306$ & $-545.47467$ \\ 
 &  & $+  6.2831853 i$ & $+  33.35282 i$ & $+  57.516544 i$ & $-73.412472 i$ \\ 
$J^{D}_{ 4}$ & $        0$ & $        0$ & $-9.3465893$ & $-122.24172$ & $-894.31476$ \\ 
$J^{D}_{ 5}$ & $   -1.125$ & $-10.113215$ & $-34.840045$ & $-15.424624$ & $ 354.88999$ \\ 
 &  & $-4.712389 i$ & $-52.488401 i$ & $-282.51753 i$ & $-923.44496 i$ \\ 
 \hline 
\end{tabular}
}
\end{center}
\caption{
Numerical results for the first five terms of the $\eps$-expansion of the master integrals $J^D_{1}$-$J^D_{5}$ 
for the kinematic point specified by eq.~(\ref{def_kinematic_point}).
}
\label{table_numerical_results_topo_D}
\end{table}
\begin{table}[!htbp]
\begin{center}
{\small
\begin{tabular}{|l|lllll|}
 \hline 
 & $\eps^0$ & $\eps^1$ & $\eps^2$ & $\eps^3$ & $\eps^4$ \\
 \hline 
$J^{E}_{ 1}$ & $        1$ & $-17.721808$ & $ 161.96604$ & $-1018.2849$ & $ 4958.5733$ \\ 
$J^{E}_{ 2}$ & $       -1$ & $ 17.721806$ & $-161.96601$ & $ 1018.2846$ & $-4958.5714$ \\ 
$J^{E}_{ 3}$ & $        1$ & $-17.721806$ & $ 161.96601$ & $-1018.2846$ & $ 4958.5715$ \\ 
$J^{E}_{ 4}$ & $        1$ & $ 6.2831853 i$ & $-18.094275$ & $-10.417826$ & $ 21.375884$ \\ 
 &  &  &  & $-31.006277 i$ & $-65.457134 i$ \\ 
$J^{E}_{ 5}$ & $        0$ & $ 0.002941742$ & $-0.052132982$ & $ 0.47646222$ & $-2.9955308$ \\ 
$J^{E}_{ 6}$ & $    -0.25$ & $ 5.4086524e-07$ & $ 4.5221719$ & $ 2.6153956$ & $-5.4032243$ \\ 
 &  & $-1.5707963 i$ & $-0.00044213247 i$ & $+  7.7502428 i$ & $+  16.362486 i$ \\ 
$J^{E}_{ 7}$ & $        0$ & $        0$ & $ 0.0013984474$ & $-0.01094393$ & $ 0.059232976$ \\ 
 &  &  & $+  0.00044553083 i$ & $+  0.0013366082 i$ & $+  0.0018112415 i$ \\ 
$J^{E}_{ 8}$ & $        0$ & $ 2.7043262e-07$ & $-3.9812567e-06$ & $ 3.0234529e-05$ & $-0.00015754108$ \\ 
$J^{E}_{ 9}$ & $    -0.25$ & $-1.5707963 i$ & $ 4.5221702$ & $ 2.6154006$ & $-5.4032039$ \\ 
 &  &  & $-0.00044553083 i$ & $+  7.7502326 i$ & $+  16.362472 i$ \\ 
$J^{E}_{ 10}$ & $        0$ & $        0$ & $ 0.00015520336$ & $-0.0010587544$ & $ 0.004771572$ \\ 
 &  &  & $+  5.5026642e-05 i$ & $+  0.00016508381 i$ & $+  0.00022371167 i$ \\ 
 \hline 
\end{tabular}
}
\end{center}
\caption{
Numerical results for the first five terms of the $\eps$-expansion of the master integrals $J^E_{1}$-$J^E_{10}$ 
for the kinematic point specified by eq.~(\ref{def_kinematic_point}).
}
\label{table_numerical_results_topo_E}
\end{table}
\begin{table}[!htbp]
\begin{center}
{\small
\begin{tabular}{|l|lllll|}
 \hline 
 & $\eps^0$ & $\eps^1$ & $\eps^2$ & $\eps^3$ & $\eps^4$ \\
 \hline 
$J^{F}_{ 1}$ & $        1$ & $ 11.138385$ & $ 60.386872$ & $ 199.16752$ & $ 381.04712$ \\ 
$J^{F}_{ 2}$ & $        1$ & $-17.721806$ & $ 161.96601$ & $-1018.2846$ & $ 4958.5715$ \\ 
$J^{F}_{ 3}$ & $        1$ & $ 6.2831853 i$ & $-18.094275$ & $-10.417826$ & $ 21.375884$ \\ 
 &  &  &  & $-31.006277 i$ & $-65.457134 i$ \\ 
$J^{F}_{ 4}$ & $        1$ & $ 39.998578$ & $-41.192268$ & $  1416.62$ & $-4196.4777$ \\ 
$J^{F}_{ 5}$ & $    -2.25$ & $-16.707577$ & $-30.485748$ & $ 195.29664$ & $ 1596.8385$ \\ 
 &  & $-4.712389 i$ & $-52.488401 i$ & $-284.3299 i$ & $-937.45111 i$ \\ 
$J^{F}_{ 6}$ & $    -0.25$ & $-1.5707963 i$ & $ 4.5221702$ & $ 2.6154006$ & $-5.4032039$ \\ 
 &  &  & $-0.00044553083 i$ & $+  7.7502326 i$ & $+  16.362472 i$ \\ 
$J^{F}_{ 7}$ & $    -1.25$ & $-11.138386$ & $-24.846011$ & $ 127.58233$ & $ 1069.9622$ \\ 
 &  & $-1.5707963 i$ & $-34.991825 i$ & $-197.30353 i$ & $-641.33006 i$ \\ 
 \hline 
\end{tabular}
}
\end{center}
\caption{
Numerical results for the first five terms of the $\eps$-expansion of the master integrals $J^F_{1}$-$J^F_{7}$ 
for the kinematic point specified by eq.~(\ref{def_kinematic_point}).
}
\label{table_numerical_results_topo_F}
\end{table}
\begin{table}[!htbp]
\begin{center}
{\small
\begin{tabular}{|l|lllll|}
 \hline 
 & $\eps^0$ & $\eps^1$ & $\eps^2$ & $\eps^3$ & $\eps^4$ \\
 \hline 
$J^{G}_{ 1}$ & $        1$ & $-17.721808$ & $ 161.96604$ & $-1018.2849$ & $ 4958.5733$ \\ 
$J^{G}_{ 2}$ & $       -1$ & $ 17.721806$ & $-161.96601$ & $ 1018.2846$ & $-4958.5714$ \\ 
$J^{G}_{ 3}$ & $        1$ & $ 6.2831853 i$ & $-18.094275$ & $-10.417826$ & $ 21.375884$ \\ 
 &  &  &  & $-31.006277 i$ & $-65.457134 i$ \\ 
$J^{G}_{ 4}$ & $    -0.25$ & $ 5.4086524e-07$ & $ 4.5221719$ & $ 2.6153956$ & $-5.4032243$ \\ 
 &  & $-1.5707963 i$ & $-0.00044213247 i$ & $+  7.7502428 i$ & $+  16.362486 i$ \\ 
$J^{G}_{ 5}$ & $        0$ & $        0$ & $ 0.0013984474$ & $-0.01094393$ & $ 0.059232976$ \\ 
 &  &  & $+  0.00044553083 i$ & $+  0.0013366082 i$ & $+  0.0018112415 i$ \\ 
 \hline 
\end{tabular}
}
\end{center}
\caption{
Numerical results for the first five terms of the $\eps$-expansion of the master integrals $J^G_{1}$-$J^G_{5}$ 
for the kinematic point specified by eq.~(\ref{def_kinematic_point}).
}
\label{table_numerical_results_topo_G}
\end{table}
\begin{table}[!htbp]
\begin{center}
{\small
\begin{tabular}{|l|lllll|}
 \hline 
 & $\eps^0$ & $\eps^1$ & $\eps^2$ & $\eps^3$ & $\eps^4$ \\
 \hline 
$J^{H}_{ 1}$ & $        1$ & $ 11.138385$ & $ 60.386872$ & $ 199.16752$ & $ 381.04712$ \\ 
$J^{H}_{ 2}$ & $        1$ & $ 6.2831853 i$ & $-18.094275$ & $-10.417826$ & $ 21.375884$ \\ 
 &  &  &  & $-31.006277 i$ & $-65.457134 i$ \\ 
$J^{H}_{ 3}$ & $    -2.25$ & $-16.707577$ & $-30.485748$ & $ 195.29664$ & $ 1596.8385$ \\ 
 &  & $-4.712389 i$ & $-52.488401 i$ & $-284.3299 i$ & $-937.45111 i$ \\ 
 \hline 
\end{tabular}
}
\end{center}
\caption{
Numerical results for the first five terms of the $\eps$-expansion of the master integrals $J^H_{1}$-$J^H_{3}$ 
for the kinematic point specified by eq.~(\ref{def_kinematic_point}).
}
\label{table_numerical_results_topo_H}
\end{table}

\subsection{Large logarithms}
\label{sect:large_logs}

In the kinematic region of interest (i.e. small $(-t)$ and large $m_Z^2$)
\bq
 \ln\left(\frac{-t}{s}\right)
 & \mbox{and} &
 \ln\left(\frac{s}{m_Z^2}\right)
\eq
are large logarithms.
Below we list for all master integrals the leading logarithms.
At order $\eps^j$ we can have at most $j$ powers of large logarithms.
The leading logarithms are the ones which occur to power $j$ at order $\eps^j$.
We remark that this counting defines at order $\eps^0$ constants as leading logarithms.
In some topologies we use different variables.
Since we have for small values of $(-t)$ 
\bq
 -t \; \sim \; \bar{y}^2 \; \sim \; \tilde{y}^2
\eq
and since we have for large values of $m_Z^2$
\bq
 m_Z^2 \; \sim \; z^{-1} \; \sim \; \tilde{z}^{-2}
\eq
the discussion carries over in a straightforward way to the new variables.
Let us stress that the following formulae are provided only to quickly gauge the numerical importance
of all master integrals. The analytic results are exact and can be used to extract not only
the leading logarithms, but also all sub-leading ones as well as the non-logarithmic part.
We use the following notation:
\bq
 L_t \; = \; \ln\left(\frac{-t}{s}\right),
 \;\;\;
 L_Z \; = \; \ln\left(\frac{s}{m_Z^2}\right),
 \;\;\;
 L_{\bar{y}} \; = \; \ln\left(\bar{y}\right),
 \;\;\;
 L_{z} \; = \; \ln\left(z\right),
 \;\;\;
 L_{\tilde{z}} \; = \; \ln\left(\tilde{z}\right).
\eq
Comparing the expressions for the leading logarithms with the numerical results from section~\ref{sect:numerical_results}
we see -- as expected -- a correlation between large logarithms and large numerical values.

\subsubsection{Topology A}

\bq
 J^{A}_{1}
 & = &
 1 + 2 L_z \eps + 2 L_z^2 \eps^2 + \frac{4}{3} L_z^3 \eps^3 + \frac{2}{3} L_z^4 \eps^4 
 + \mathrm{subleading} + {\mathcal O}\left(\eps^5\right),
 \nonumber \\
 J^{A}_{2}
 & = &
 -1 - 2 L_z \eps - 2 L_z^2 \eps^2 - \frac{4}{3} L_z^3 \eps^3 - \frac{2}{3} L_z^4 \eps^4 
 + \mathrm{subleading} + {\mathcal O}\left(\eps^5\right),
 \nonumber \\
 J^{A}_{3}
 & = &
 \frac{3}{2}
 + \mathrm{subleading} + {\mathcal O}\left(\eps^5\right),
 \nonumber \\
 J^{A}_{4}
 & = &
 1 + 2 L_z \eps + 2 L_z^2 \eps^2 + \frac{4}{3} L_z^3 \eps^3 + \frac{2}{3} L_z^4 \eps^4 
 + \mathrm{subleading} + {\mathcal O}\left(\eps^5\right),
 \nonumber \\
 J^{A}_{5}
 & = &
 \frac{1}{2}
 + \mathrm{subleading} + {\mathcal O}\left(\eps^5\right),
 \nonumber \\
 J^{A}_{6}
 & = &
 0
 + \mathrm{subleading} + {\mathcal O}\left(\eps^5\right),
 \nonumber \\
 J^{A}_{7}
 & = &
 2 L_z \eps + 2 L_z^2 \eps^2 + \frac{4}{3} L_z^3 \eps^3 + \frac{2}{3} L_z^4 \eps^4 
 + \mathrm{subleading} + {\mathcal O}\left(\eps^5\right),
 \nonumber \\
 J^{A}_{8}
 & = &
 0
 + \mathrm{subleading} + {\mathcal O}\left(\eps^5\right),
 \nonumber \\
 J^{A}_{9}
 & = &
 -2 L_z \eps - 2 L_z^2 \eps^2 - \frac{4}{3} L_z^3 \eps^3 - \frac{2}{3} L_z^4 \eps^4 
 + \mathrm{subleading} + {\mathcal O}\left(\eps^5\right),
 \nonumber \\
 J^{A}_{10}
 & = &
 - \frac{1}{8}
 + \mathrm{subleading} + {\mathcal O}\left(\eps^5\right),
 \nonumber \\
 J^{A}_{11}
 & = &
 0
 + \mathrm{subleading} + {\mathcal O}\left(\eps^5\right),
 \nonumber \\
 J^{A}_{12}
 & = &
 0
 + \mathrm{subleading} + {\mathcal O}\left(\eps^5\right),
 \nonumber \\
 J^{A}_{13}
 & = &
 - \frac{1}{8}
 + \mathrm{subleading} + {\mathcal O}\left(\eps^5\right),
 \nonumber \\
 J^{A}_{14}
 & = &
 0
 + \mathrm{subleading} + {\mathcal O}\left(\eps^5\right),
 \nonumber \\
 J^{A}_{15}
 & = &
 0
 + \mathrm{subleading} + {\mathcal O}\left(\eps^5\right).
\eq

\subsubsection{Topology B}

\bq
 J^{B}_{1}
 & = &
 1 -4 L_{\bar{y}} \eps + 8 L_{\bar{y}}^2 \eps^2 - \frac{32}{3} L_{\bar{y}}^3 \eps^3 + \frac{32}{3} L_{\bar{y}}^4 \eps^4 
 + \mathrm{subleading} + {\mathcal O}\left(\eps^5\right),
 \nonumber \\
 J^{B}_{2}
 & = &
 \frac{3}{2}
 + \mathrm{subleading} + {\mathcal O}\left(\eps^5\right),
 \nonumber \\
 J^{B}_{3}
 & = &
 1 + 2 L_z \eps + 2 L_z^2 \eps^2 + \frac{4}{3} L_z^3 \eps^3 + \frac{2}{3} L_z^4 \eps^4 
 + \mathrm{subleading} + {\mathcal O}\left(\eps^5\right),
 \nonumber \\
 J^{B}_{4}
 & = &
 \frac{1}{2}
 + \mathrm{subleading} + {\mathcal O}\left(\eps^5\right),
 \nonumber \\
 J^{B}_{5}
 & = &
 0
 + \mathrm{subleading} + {\mathcal O}\left(\eps^5\right),
 \nonumber \\
 J^{B}_{6}
 & = &
 1 - \left(2L_z+8L_{\bar{y}}\right) \eps - \left(2L_z^2 -16L_{\bar{y}}^2\right)\eps^2 - \left( \frac{4}{3}L_z^3 + \frac{64}{3}L_{\bar{y}}^3\right) \eps^3
 - \left( \frac{2}{3}L_z^4 - \frac{64}{3} L_{\bar{y}}^4 \right) \eps^4
 \nonumber \\
 & &  
 + \mathrm{subleading} + {\mathcal O}\left(\eps^5\right),
 \nonumber \\
 J^{B}_{7}
 & = &
 -2 L_z \eps - 2 L_z^2 \eps^2 - \frac{4}{3} L_z^3 \eps^3 - \frac{2}{3} L_z^4 \eps^4
 + \mathrm{subleading} + {\mathcal O}\left(\eps^5\right),
 \nonumber \\
 J^{B}_{8}
 & = &
 - \frac{9}{8} + 3 L_{\bar{y}} \eps - 3 L_{\bar{y}}^2 \eps^2 + 4 L_{\bar{y}}^4 \eps^4
 + \mathrm{subleading} + {\mathcal O}\left(\eps^5\right),
 \nonumber \\
 J^{B}_{9}
 & = &
 -\frac{1}{8}
 + \mathrm{subleading} + {\mathcal O}\left(\eps^5\right),
 \nonumber \\
 J^{B}_{10}
 & = &
 0
 + \mathrm{subleading} + {\mathcal O}\left(\eps^5\right),
 \nonumber \\
 J^{B}_{11}
 & = &
 - \frac{5}{8} + 2 L_{\bar{y}} \eps -2 L_{\bar{y}}^2 \eps^2 + \frac{8}{3} L_{\bar{y}}^4 \eps^4
 + \mathrm{subleading} + {\mathcal O}\left(\eps^5\right).
\eq

\subsubsection{Topology C}
 
\bq
 J^{C}_{1}
 & = &
 1 + 2 L_z \eps + 2 L_z^2 \eps^2 + \frac{4}{3} L_z^3 \eps^3 + \frac{2}{3} L_z^4 \eps^4 
 + \mathrm{subleading} + {\mathcal O}\left(\eps^5\right),
 \nonumber \\
 J^{C}_{2}
 & = &
 -1 - 2 L_z \eps - 2 L_z^2 \eps^2 - \frac{4}{3} L_z^3 \eps^3 - \frac{2}{3} L_z^4 \eps^4 
 + \mathrm{subleading} + {\mathcal O}\left(\eps^5\right),
 \nonumber \\
 J^{C}_{3}
 & = &
 \frac{3}{2}
 + \mathrm{subleading} + {\mathcal O}\left(\eps^5\right),
 \nonumber \\
 J^{C}_{4}
 & = &
 \frac{1}{2}
 + \mathrm{subleading} + {\mathcal O}\left(\eps^5\right),
 \nonumber \\
 J^{C}_{5}
 & = &
 0
 + \mathrm{subleading} + {\mathcal O}\left(\eps^5\right),
 \nonumber \\
 J^{C}_{6}
 & = &
 2 L_z \eps + 2 L_z^2 \eps^3 + \frac{4}{3} L_z^3 \eps^3 + \frac{2}{3} L_z^4 \eps^4
 + \mathrm{subleading} + {\mathcal O}\left(\eps^5\right),
 \nonumber \\
 J^{C}_{7}
 & = &
 - \frac{1}{8}
 + \mathrm{subleading} + {\mathcal O}\left(\eps^5\right),
 \nonumber \\
 J^{C}_{8}
 & = &
 0
 + \mathrm{subleading} + {\mathcal O}\left(\eps^5\right).
\eq

\subsubsection{Topology D}
 
\bq
 J^{D}_{1}
 & = &
 1 -4 L_{\bar{y}} \eps + 8 L_{\bar{y}}^2 \eps^2 - \frac{32}{3} L_{\bar{y}}^3 \eps^3 + \frac{32}{3} L_{\bar{y}}^4 \eps^4 
 + \mathrm{subleading} + {\mathcal O}\left(\eps^5\right),
 \nonumber \\
 J^{D}_{2}
 & = &
\frac{3}{2}
 + \mathrm{subleading} + {\mathcal O}\left(\eps^5\right),
 \nonumber \\
 J^{D}_{3}
 & = &
 \frac{1}{2}
 + \mathrm{subleading} + {\mathcal O}\left(\eps^5\right),
 \nonumber \\
 J^{D}_{4}
 & = &
 0
 + \mathrm{subleading} + {\mathcal O}\left(\eps^5\right),
 \nonumber \\
 J^{D}_{5}
 & = &
 - \frac{9}{8} + 3 L_{\bar{y}} \eps - 3 L_{\bar{y}}^2 \eps^2 + 4 L_{\bar{y}}^4 \eps^4
 + \mathrm{subleading} + {\mathcal O}\left(\eps^5\right).
\eq

\subsubsection{Topology E}
 
\bq
 J^{E}_{1}
 & = &
 1 + 4 L_{\tilde{z}} \eps + 8 L_{\tilde{z}}^2 \eps^2 + \frac{32}{3} L_{\tilde{z}}^3 \eps^3 + \frac{32}{3} L_{\tilde{z}}^4 \eps^4
 + \mathrm{subleading} + {\mathcal O}\left(\eps^5\right),
 \nonumber \\
 J^{E}_{2}
 & = &
 -1 - 4 L_{\tilde{z}} \eps - 8 L_{\tilde{z}}^2 \eps^2 - \frac{32}{3} L_{\tilde{z}}^3 \eps^3 - \frac{32}{3} L_{\tilde{z}}^4 \eps^4
 + \mathrm{subleading} + {\mathcal O}\left(\eps^5\right),
 \nonumber \\
 J^{E}_{3}
 & = &
 1 + 4 L_{\tilde{z}} \eps + 8 L_{\tilde{z}}^2 \eps^2 + \frac{32}{3} L_{\tilde{z}}^3 \eps^3 + \frac{32}{3} L_{\tilde{z}}^4 \eps^4
 + \mathrm{subleading} + {\mathcal O}\left(\eps^5\right),
 \nonumber \\
 J^{E}_{4}
 & = &
 1
 + \mathrm{subleading} + {\mathcal O}\left(\eps^5\right),
 \nonumber \\
 J^{E}_{5}
 & = &
 0
 + \mathrm{subleading} + {\mathcal O}\left(\eps^5\right),
 \nonumber \\
 J^{E}_{6}
 & = &
 - \frac{1}{4}
 + \mathrm{subleading} + {\mathcal O}\left(\eps^5\right),
 \nonumber \\
 J^{E}_{7}
 & = &
 0
 + \mathrm{subleading} + {\mathcal O}\left(\eps^5\right),
 \nonumber \\
 J^{E}_{8}
 & = &
 0
 + \mathrm{subleading} + {\mathcal O}\left(\eps^5\right),
 \nonumber \\
 J^{E}_{9}
 & = &
 - \frac{1}{4}
 + \mathrm{subleading} + {\mathcal O}\left(\eps^5\right),
 \nonumber \\
 J^{E}_{10}
 & = &
 0
 + \mathrm{subleading} + {\mathcal O}\left(\eps^5\right).
\eq

\subsubsection{Topology F}
 
\bq
 J^{F}_{1}
 & = &
 1 - 2 L_t \eps + 2 L_t^2 \eps^2 - \frac{4}{3} L_t^3 \eps^3 + \frac{2}{3} L_t^4 \eps^4 
 + \mathrm{subleading} + {\mathcal O}\left(\eps^5\right),
 \nonumber \\
 J^{F}_{2}
 & = &
 1 + 2 L_Z \eps + 2 L_Z^2 \eps^2 + \frac{4}{3} L_Z^3 \eps^3 + \frac{2}{3} L_Z^4 \eps^4
 + \mathrm{subleading} + {\mathcal O}\left(\eps^5\right),
 \nonumber \\
 J^{F}_{3}
 & = &
 1
 + \mathrm{subleading} + {\mathcal O}\left(\eps^5\right),
 \nonumber \\
 J^{F}_{4}
 & = &
 1 - \left(2L_Z + 4L_t\right) \eps - \left(2L_Z^2 - 4L_t^2\right) \eps^2 - \left( \frac{4}{3} L_Z^3 + \frac{8}{3} L_t^3 \right) \eps^3
 - \left( \frac{2}{3} L_Z^4 - \frac{4}{3} L_t^4 \right) \eps^4
 \nonumber \\
 & & 
 + \mathrm{subleading} + {\mathcal O}\left(\eps^5\right),
 \nonumber \\
 J^{F}_{5}
 & = &
 - \frac{9}{4} + 3 L_t \eps - \frac{3}{2} L_t^2 \eps^2 + \frac{1}{2} L_t^4 \eps^4
 + \mathrm{subleading} + {\mathcal O}\left(\eps^5\right),
 \nonumber \\
 J^{F}_{6}
 & = &
 - \frac{1}{4}
 + \mathrm{subleading} + {\mathcal O}\left(\eps^5\right),
 \nonumber \\
 J^{F}_{7}
 & = &
 - \frac{5}{4} + 2 L_t \eps - L_t^2 \eps^2 + \frac{1}{3} L_t^4 \eps^4
 + \mathrm{subleading} + {\mathcal O}\left(\eps^5\right).
\eq

\subsubsection{Topology G}
 
\bq
 J^{G}_{1}
 & = &
 1 + 2 L_Z \eps + 2 L_Z^2 \eps^2 + \frac{4}{3} L_Z^3 \eps^3 + \frac{2}{3} L_Z^4 \eps^4
 + \mathrm{subleading} + {\mathcal O}\left(\eps^5\right),
 \nonumber \\
 J^{G}_{2}
 & = &
 - 1 - 2 L_Z \eps - 2 L_Z^2 \eps^2 - \frac{4}{3} L_Z^3 \eps^3 - \frac{2}{3} L_Z^4 \eps^4
 + \mathrm{subleading} + {\mathcal O}\left(\eps^5\right),
 \nonumber \\
 J^{G}_{3}
 & = &
 1
 + \mathrm{subleading} + {\mathcal O}\left(\eps^5\right),
 \nonumber \\
 J^{G}_{4}
 & = &
 - \frac{1}{4}
 + \mathrm{subleading} + {\mathcal O}\left(\eps^5\right),
 \nonumber \\
 J^{G}_{5}
 & = &
 0
 + \mathrm{subleading} + {\mathcal O}\left(\eps^5\right).
\eq

\subsubsection{Topology H}
 
\bq
 J^{H}_{1}
 & = &
 1 - 2 L_t \eps + 2 L_t^2 \eps^2 - \frac{4}{3} L_t^3 \eps^3 + \frac{2}{3} L_t^4 \eps^4 
 + \mathrm{subleading} + {\mathcal O}\left(\eps^5\right),
 \nonumber \\
 J^{H}_{2}
 & = &
 1
 + \mathrm{subleading} + {\mathcal O}\left(\eps^5\right),
 \nonumber \\
 J^{H}_{3}
 & = &
 - \frac{9}{4} + 3 L_t \eps - \frac{3}{2} L_t^2 \eps^2 + \frac{1}{2} L_t^4 \eps^4
 + \mathrm{subleading} + {\mathcal O}\left(\eps^5\right).
\eq

\section{Conclusions}
\label{sect:conclusions}

In this paper we computed a set of two-loop Feynman integrals relevant to the Moller experiment
and to the P2 experiment.
We considered Feynman integrals, which are obtained from a box integral by the insertion of a light fermion loop.
The exchanged particles in the box integral are either photons or heavy electro-weak gauge bosons.
We considered all combinations.
By rationalising all occurring square roots we showed 
that all these Feynman integrals can be expressed in terms of multiple polylogarithms.
We organised our results such that all large logarithms are manifest.
Furthermore, we provided highly efficient numerical evaluation routines for all
master integrals in the kinematic region of interest.  

For the complete set of the two-loop electro-weak corrections there are of course more diagrams to be considered.
In particular, this includes the planar and non-planar double box integrals.
Again, one has to consider all possible combinations of photon and heavy gauge boson exchanges.
The case of the exchange of three heavy gauge bosons is particularly interesting:
We expect the planar double box integral with the exchange of three heavy gauge bosons to be associated 
with a curve of genus one and
the non-planar double box integral with the exchange of three heavy gauge bosons to be associated 
with a curve of genus two \cite{Marzucca:2023gto}.
This is an interesting project for the future.

\subsection*{Acknowledgments}

We would like to thank Jens Erler, Mikhail Gorchtein and Hubert Spiesberger  
for useful discussions and comments on the manuscript. 
This work has been supported by the Cluster of Excellence Precision Physics, Fundamental Interactions, and Structure of
Matter (PRISMA EXC 2118/1) funded by the German Research Foundation (DFG) within
the German Excellence Strategy (Project ID 390831469).

\begin{appendix}

\section{Master sectors}
\label{sect_master_sectors}

\begin{figure}[!hbp]
\begin{center}
\includegraphics[scale=1.0]{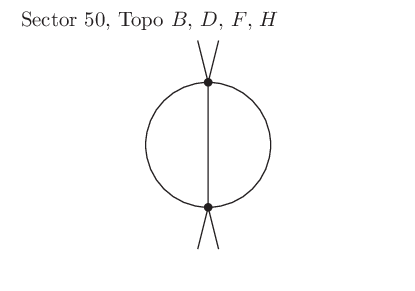}
\includegraphics[scale=1.0]{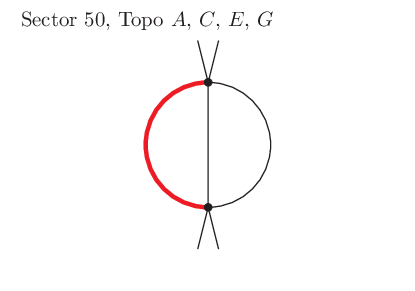}
\includegraphics[scale=1.0]{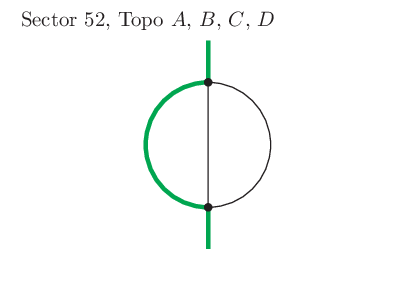}
\includegraphics[scale=1.0]{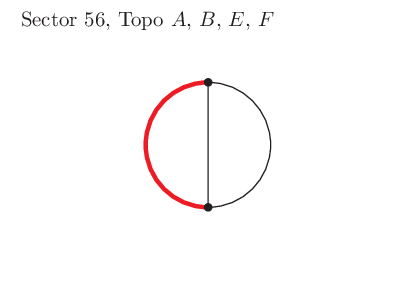}
\includegraphics[scale=1.0]{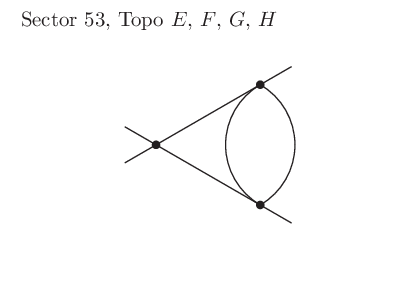}
\includegraphics[scale=1.0]{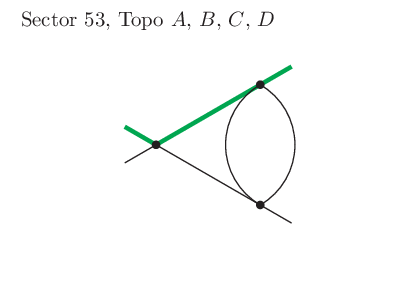}
\includegraphics[scale=1.0]{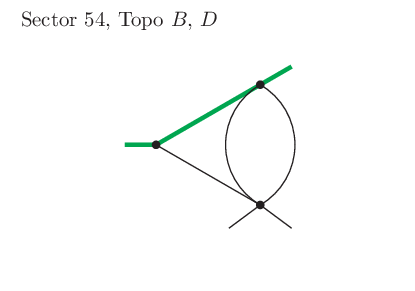}
\includegraphics[scale=1.0]{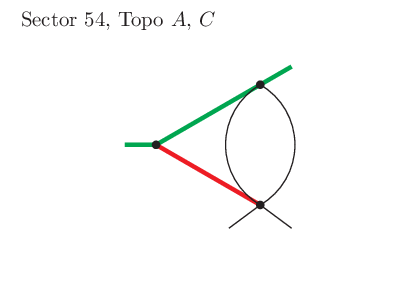}
\end{center}
\caption{
Master sectors (part 1).
}
\label{fig_master_topologies_part1}
\end{figure}

\begin{figure}
\begin{center}
\includegraphics[scale=1.0]{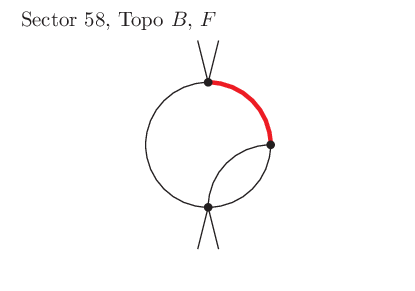}
\includegraphics[scale=1.0]{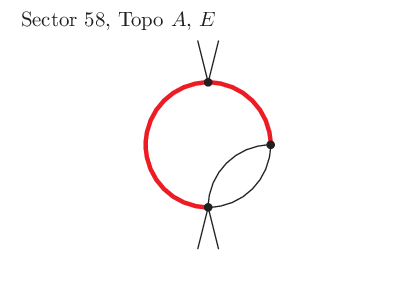}
\includegraphics[scale=1.0]{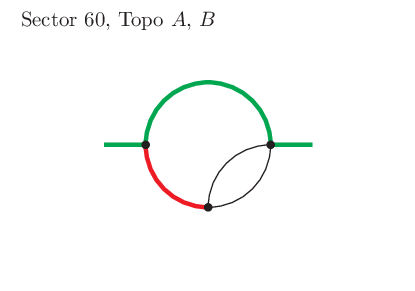}
\includegraphics[scale=1.0]{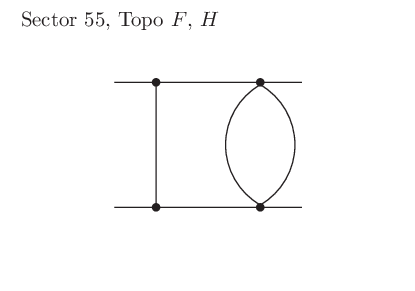}
\includegraphics[scale=1.0]{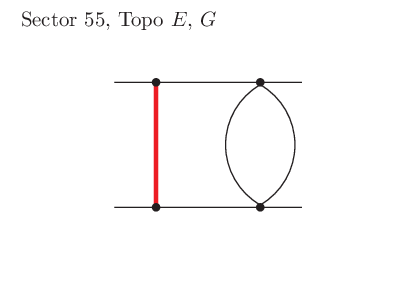}
\includegraphics[scale=1.0]{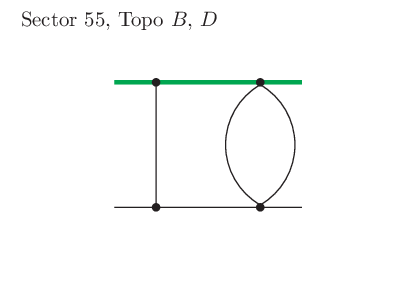}
\includegraphics[scale=1.0]{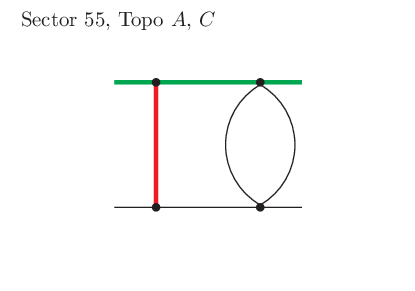}
\includegraphics[scale=1.0]{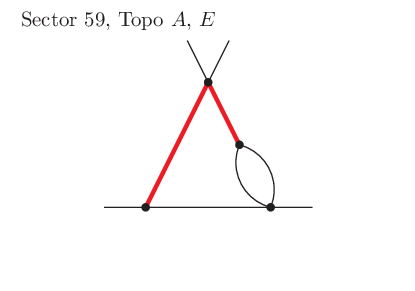}
\end{center}
\caption{
Master sectors (part 2).
}
\label{fig_master_topologies_part2}
\end{figure}

\begin{figure}
\begin{center}
\includegraphics[scale=1.0]{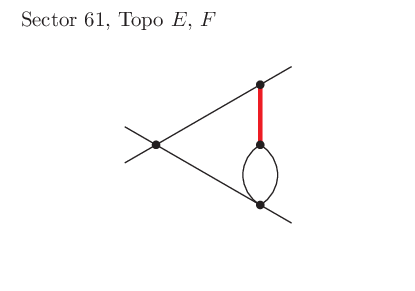}
\includegraphics[scale=1.0]{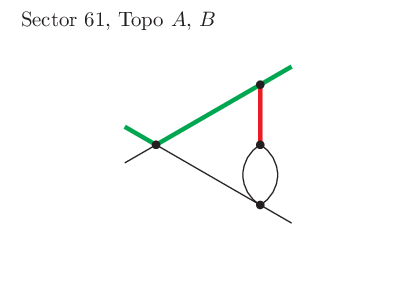}
\includegraphics[scale=1.0]{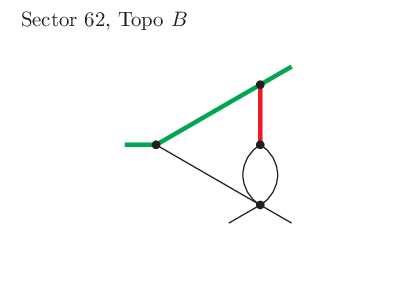}
\includegraphics[scale=1.0]{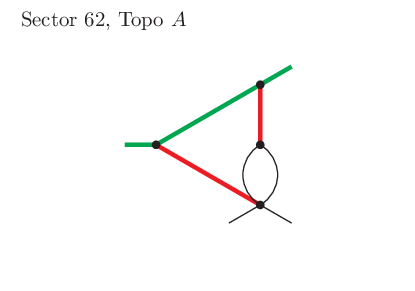}
\includegraphics[scale=1.0]{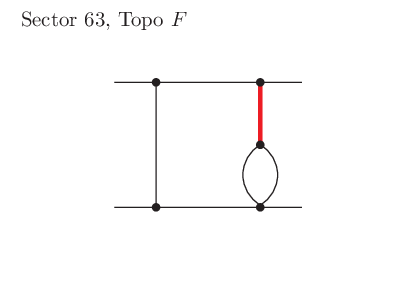}
\includegraphics[scale=1.0]{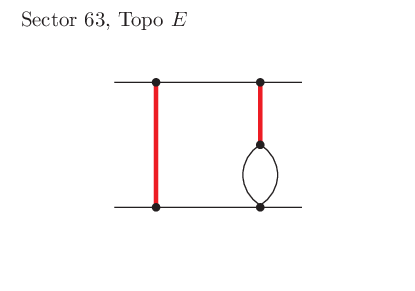}
\includegraphics[scale=1.0]{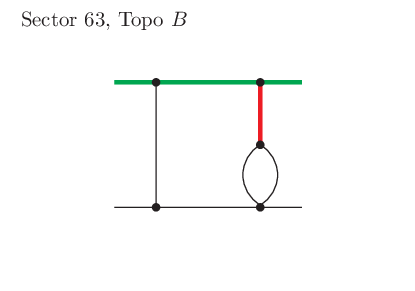}
\includegraphics[scale=1.0]{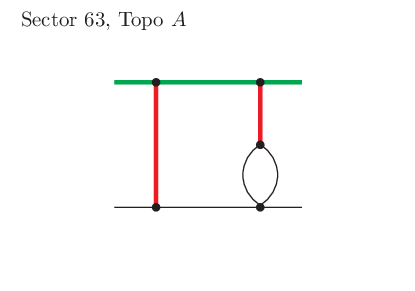}
\end{center}
\caption{
Master sectors (part 3).
}
\label{fig_master_topologies_part3}
\end{figure}

In this appendix we show in tables~\ref{fig_master_topologies_part1}-\ref{fig_master_topologies_part3} 
the diagrams of all master sectors.
The colour coding is as follows: A green line indicates a particle of mass $m_N$, a red line a particle of mass $m_Z$.
Uncoloured lines are massless.

\section{Supplementary material}
\label{sect:supplement}

Attached to the arxiv version of this article are for each topology
\bq
 X & \in & \left\{ A,B,C,D,E,F,G,H \right\}
\eq
the electronic files
\begin{center}
 \verb|topo_X_symbolic.mpl|, \; \verb|topo_X_numeric.cc|.
\end{center}
The first file is in {\tt Maple} syntax and 
defines the transformation matrix $U^X$ appearing in eq.~(\ref{def_fibre_transformation}),
its inverse $(U^X)^{-1}$ and the matrix $M^X$ appearing in the differential equation~(\ref{def_differential_equation}).
These are denoted as 
\begin{center}
 \verb|U_X|, \; \verb|Uinv_X|, \; \verb|M_X|.
\end{center}
The second file \verb|topo_X_numeric.cc| is a {\tt C++}-program and provides numerical evaluation routines
for all master integrals of a given topology.
This {\tt C++}-program requires the \verb|GiNaC|-library \cite{Bauer:2000cp}.

\section{One-loop integrals}
\label{sect_one_loop}

For convenience we also include the corresponding one-loop integrals.
At one-loop we consider the family of integrals
\bq
\label{def_integral_one_loop}
 I_{\nu_1 \nu_2 \nu_3 \nu_4}
 & = &
 e^{\Eulerconstant \eps}
 \left(\mu^2\right)^{\nu-\frac{D}{2}}
 \int \frac{d^Dk_1}{i \pi^{\frac{D}{2}}} 
 \prod\limits_{j=1}^4 \frac{1}{ P_j^{\nu_j} },
\eq
where the notation is as in section~\ref{sect:preliminaries}.
At one-loop, the topologies $B$ and $C$ are related by symmetry, and so are the topologies $F$ and $G$.
We therefore have to consider only the topologies $A$, $B$, $D$, $E$, $F$ and $H$.
Possible pre-canonical bases are:
\bq
 K^A & = &
 \left(
  I^A_{0100},
  I^A_{0010},
  I^A_{1010},
  I^A_{0110},
  I^A_{0101},
  I^A_{1110}
  I^A_{1101}
  I^A_{0111},
  I^A_{1111}
 \right)^T,
 \nonumber \\
 K^B & = &
 \left(
  I^B_{0010},
  I^B_{0001},
  I^B_{1010},
  I^B_{0101},
  I^B_{0011},
  I^B_{1011}
  I^B_{0111},
  I^B_{1111}
 \right)^T,
 \nonumber \\
 K^D & = &
 \left(
  I^D_{0010},
  I^D_{1010},
  I^D_{0101},
  I^D_{0111},
  I^D_{1111}
 \right)^T,
 \nonumber \\
 K^E & = &
 \left(
  I^E_{0100},
  I^E_{1010},
  I^E_{0101},
  I^E_{1110}
  I^E_{1101},
  I^E_{1111}
 \right)^T,
 \nonumber \\
 K^F & = &
 \left(
  I^F_{0001},
  I^F_{1010},
  I^F_{0101},
  I^F_{1011}
  I^F_{1111}
 \right)^T,
 \nonumber \\
 K^H & = &
 \left(
  I^H_{1010},
  I^H_{0101},
  I^H_{1111}
 \right)^T.
\eq
Below we list a possible choice of master integrals of uniform transcendental weight, together with the corresponding leading logarithms.
For the logarithms we use the same notation as in section~\ref{sect:large_logs}.

\subsubsection*{Topology A}

\begin{alignat}{3}
 \mbox{Sector 2:} \;\;\;\; &
 L^{A}_{1}
 & = \;\; & 
 \eps
 \; {\bf D}^- I^A_{0100}
 & \; = \; &
 1 + L_z \eps + \frac{1}{2} L_z^2 \eps^2
 + \mathrm{subleading} + {\mathcal O}\left(\eps^3\right),
 \nonumber \\
 \mbox{Sector 4:} \;\;\;\; &
 L^{A}_{2}
 & = \;\; & 
 \eps
 \; {\bf D}^- I^A_{0010}
 & \; = \; &
 1
 + \mathrm{subleading} + {\mathcal O}\left(\eps^3\right),
 \nonumber \\
 \mbox{Sector 5:} \;\;\;\; &
 L^{A}_{3}
 & = \;\; & 
 \eps
 \left(\frac{m_N^2-s}{\mu^2}\right)
 \; {\bf D}^- I^A_{1010}
 & \; = \; &
 -1
 + \mathrm{subleading} + {\mathcal O}\left(\eps^3\right),
 \nonumber \\
 \mbox{Sector 6:} \;\;\;\; &
 L^{A}_{4}
 & = \;\; & 
 \eps 
 \left(\frac{r_3}{\mu^2}\right)
 \; {\bf D}^- I^A_{0110}
 & \; = \; &
 - L_z \eps - \frac{1}{2} L_z^2 \eps^2
 + \mathrm{subleading} + {\mathcal O}\left(\eps^3\right),
 \nonumber \\
 \mbox{Sector 10:} \;\;\;\; &
 L^{A}_{5}
 & = \;\; & 
 \eps 
 \left(\frac{r_2}{\mu^2}\right)
 \; {\bf D}^- I^A_{0101}
 & \; = \; &
 0
 + \mathrm{subleading} + {\mathcal O}\left(\eps^3\right),
 \nonumber \\
 \mbox{Sector 7:} \;\;\;\; &
 L^{A}_{6}
 & = \;\; & 
 \eps^2
 \left(\frac{m_N^2-s}{\mu^2}\right)
 \; I^A_{1110}
 & \; = \; &
 0
 + \mathrm{subleading} + {\mathcal O}\left(\eps^3\right),
 \nonumber \\
 \mbox{Sector 11:} \;\;\;\; &
 L^{A}_{7}
 & = \;\; & 
 \eps^2
 \left(\frac{-t}{\mu^2}\right)
 \; I^A_{1101}
 & \; = \; &
 0
 + \mathrm{subleading} + {\mathcal O}\left(\eps^3\right),
 \nonumber \\
 \mbox{Sector 14:} \;\;\;\; &
 L^{A}_{8}
 & = \;\; & 
 \eps^2
 \left(\frac{r_1}{\mu^2}\right)
 \; I^A_{0111}
 & \; = \; &
 0
 + \mathrm{subleading} + {\mathcal O}\left(\eps^3\right),
 \nonumber \\
 \mbox{Sector 15:} \;\;\;\; &
 L^{A}_{9}
 & = \;\; & 
 \eps^2 
 \left(\frac{r_5}{\mu^4}\right)
 \; I^A_{1111}
 & \; = \; &
 0
 + \mathrm{subleading} + {\mathcal O}\left(\eps^3\right).
\end{alignat}

\subsubsection*{Topology B}

\begin{alignat}{3}
 \mbox{Sector 4:} \;\;\;\; &
 L^{B}_{1}
 & = \;\; & 
 \eps
 \; {\bf D}^- I^B_{0010}
 & \; = \; &
 1
 + \mathrm{subleading} + {\mathcal O}\left(\eps^3\right),
 \nonumber \\
 \mbox{Sector 8:} \;\;\;\; &
 L^{B}_{2}
 & = \;\; & 
 \eps
 \; {\bf D}^- I^B_{0001}
 & \; = \; &
 1 + L_z \eps + \frac{1}{2} L_z^2 \eps^2
 + \mathrm{subleading} + {\mathcal O}\left(\eps^3\right),
 \nonumber \\
 \mbox{Sector 5:} \;\;\;\; &
 L^{B}_{3}
 & = \;\; & 
 \eps
 \left(\frac{m_N^2-s}{\mu^2}\right)
 \; {\bf D}^- I^B_{1010}
 & \; = \; &
 -1
 + \mathrm{subleading} + {\mathcal O}\left(\eps^3\right),
 \nonumber \\
 \mbox{Sector 10:} \;\;\;\; &
 L^{B}_{4}
 & = \;\; & 
 \eps 
 \left(\frac{m_Z^2-t}{\mu^2}\right)
 \; {\bf D}^- I^B_{0101}
 & \; = \; &
 -1 - L_z \eps - \frac{1}{2} L_z^2 \eps^2
 + \mathrm{subleading} + {\mathcal O}\left(\eps^3\right),
 \nonumber \\
 \mbox{Sector 12:} \;\;\;\; &
 L^{B}_{5}
 & = \;\; & 
 \eps 
 \left(\frac{r_3}{\mu^2}\right)
 \; {\bf D}^- I^B_{0011}
 & \; = \; &
 - L_z \eps - \frac{1}{2} L_z^2 \eps^2
 + \mathrm{subleading} + {\mathcal O}\left(\eps^3\right),
 \nonumber \\
 \mbox{Sector 13:} \;\;\;\; &
 L^{B}_{6}
 & = \;\; & 
 \eps^2
 \left(\frac{m_N^2-s}{\mu^2}\right)
 \; I^B_{1011}
 & \; = \; &
 0
 + \mathrm{subleading} + {\mathcal O}\left(\eps^3\right),
 \nonumber \\
 \mbox{Sector 14:} \;\;\;\; &
 L^{B}_{7}
 & = \;\; & 
 \eps^2
 \left(\frac{r_1}{\mu^2}\right)
 \; I^B_{0111}
 & \; = \; &
 0
 + \mathrm{subleading} + {\mathcal O}\left(\eps^3\right),
 \nonumber \\
 \mbox{Sector 15:} \;\;\;\; &
 L^{B}_{8}
 & = \;\; & 
 \eps^2 
 \left(\frac{m_N^2-s}{\mu^2}\right)
 \left(\frac{m_Z^2-t}{\mu^2}\right)
 \; I^B_{1111}
 & \; = \; &
 \frac{1}{2}
 + \mathrm{subleading} + {\mathcal O}\left(\eps^3\right).
\end{alignat}

\subsubsection*{Topology D}

\begin{alignat}{3}
 \mbox{Sector 4:} \;\;\;\; &
 L^{D}_{1}
 & = \;\; & 
 \eps
 \; {\bf D}^- I^D_{0010}
 & \; = \; &
 1
 + \mathrm{subleading} + {\mathcal O}\left(\eps^3\right),
 \nonumber \\
 \mbox{Sector 5:} \;\;\;\; &
 L^{D}_{2}
 & = \;\; & 
 \eps
 \left(\frac{m_N^2-s}{\mu^2}\right)
 \; {\bf D}^- I^D_{1010}
 & \; = \; &
 -1
 + \mathrm{subleading} + {\mathcal O}\left(\eps^3\right),
 \nonumber \\
 \mbox{Sector 10:} \;\;\;\; &
 L^{D}_{3}
 & = \;\; & 
 \eps 
 \left(\frac{-t}{\mu^2}\right)
 \; {\bf D}^- I^D_{0101}
 & \; = \; &
 -2 + 4 L_{\bar{y}} \eps - 4 L_{\bar{y}}^2 \eps^2
 + \mathrm{subleading} + {\mathcal O}\left(\eps^3\right),
 \nonumber \\
 \mbox{Sector 14:} \;\;\;\; &
 L^{D}_{4}
 & = \;\; & 
 \eps^2
 \left(\frac{r_1}{\mu^2}\right)
 \; I^D_{0111}
 & \; = \; &
 0
 + \mathrm{subleading} + {\mathcal O}\left(\eps^3\right),
 \nonumber \\
 \mbox{Sector 15:} \;\;\;\; &
 L^{D}_{5}
 & = \;\; & 
 \eps^2 
 \left(\frac{m_N^2-s}{\mu^2}\right)
 \left(\frac{-t}{\mu^2}\right)
 \; I^D_{1111}
 & \; = \; &
 2 - 2 L_{\bar{y}} \eps
 + \mathrm{subleading} + {\mathcal O}\left(\eps^3\right).
\end{alignat}

\subsubsection*{Topology E}

\begin{alignat}{3}
 \mbox{Sector 2:} \;\;\;\; &
 L^{E}_{1}
 & = \;\; & 
 \eps
 \; {\bf D}^- I^E_{0100}
 & \; = \; &
 1 + 2 L_{\tilde{z}} \eps + 2 L_{\tilde{z}}^2 \eps^2
 + \mathrm{subleading} + {\mathcal O}\left(\eps^3\right),
 \nonumber \\
 \mbox{Sector 5:} \;\;\;\; &
 L^{E}_{2}
 & = \;\; & 
 \eps
 \left(\frac{-s}{\mu^2}\right)
 \; {\bf D}^- I^E_{1010}
 & \; = \; &
 -2
 + \mathrm{subleading} + {\mathcal O}\left(\eps^3\right),
 \nonumber \\
 \mbox{Sector 10:} \;\;\;\; &
 L^{E}_{3}
 & = \;\; & 
 \eps 
 \left(\frac{r_2}{\mu^2}\right)
 \; {\bf D}^- I^E_{0101}
 & \; = \; &
 0
 + \mathrm{subleading} + {\mathcal O}\left(\eps^3\right),
 \nonumber \\
 \mbox{Sector 7:} \;\;\;\; &
 L^{E}_{4}
 & = \;\; & 
 \eps^2
 \left(\frac{-s}{\mu^2}\right)
 \; I^E_{1110}
 & \; = \; &
 0
 + \mathrm{subleading} + {\mathcal O}\left(\eps^3\right),
 \nonumber \\
 \mbox{Sector 11:} \;\;\;\; &
 L^{E}_{5}
 & = \;\; & 
 \eps^2
 \left(\frac{-t}{\mu^2}\right)
 \; I^E_{1101}
 & \; = \; &
 0
 + \mathrm{subleading} + {\mathcal O}\left(\eps^3\right),
 \nonumber \\
 \mbox{Sector 15:} \;\;\;\; &
 L^{E}_{6}
 & = \;\; & 
 \eps^2 
 \left(\frac{r_4}{\mu^4}\right)
 \; I^E_{1111}
 & \; = \; &
 0
 + \mathrm{subleading} + {\mathcal O}\left(\eps^3\right).
\end{alignat}

\subsubsection*{Topology F}

\begin{alignat}{3}
 \mbox{Sector 8:} \;\;\;\; &
 L^{F}_{1}
 & = \;\; & 
 \eps
 \; {\bf D}^- I^F_{0001}
 & \; = \; &
 1+ L_Z \eps + \frac{1}{2} L_Z^2 \eps^2
 + \mathrm{subleading} + {\mathcal O}\left(\eps^3\right),
 \nonumber \\
 \mbox{Sector 5:} \;\;\;\; &
 L^{F}_{2}
 & = \;\; & 
 \eps
 \left(\frac{-s}{\mu^2}\right)
 \; {\bf D}^- I^F_{1010}
 & \; = \; &
 -2
 + \mathrm{subleading} + {\mathcal O}\left(\eps^3\right),
 \nonumber \\
 \mbox{Sector 10:} \;\;\;\; &
 L^{F}_{3}
 & = \;\; & 
 \eps 
 \left(\frac{m_Z^2-t}{\mu^2}\right)
 \; {\bf D}^- I^F_{0101}
 & \; = \; &
 -1 - L_Z \eps - \frac{1}{2} L_Z^2 \eps^2
 + \mathrm{subleading} + {\mathcal O}\left(\eps^3\right),
 \nonumber \\
 \mbox{Sector 13:} \;\;\;\; &
 L^{F}_{4}
 & = \;\; & 
 \eps^2
 \left(\frac{-s}{\mu^2}\right)
 \; I^F_{1011}
 & \; = \; &
 0
 + \mathrm{subleading} + {\mathcal O}\left(\eps^3\right),
 \nonumber \\
 \mbox{Sector 15:} \;\;\;\; &
 L^{F}_{5}
 & = \;\; & 
 \eps^2 
 \left(\frac{-s}{\mu^2}\right)
 \left(\frac{m_Z^2-t}{\mu^2}\right)
 \; I^F_{1111}
 & \; = \; &
 1
 + \mathrm{subleading} + {\mathcal O}\left(\eps^3\right).
\end{alignat}

\subsubsection*{Topology H}

\begin{alignat}{3}
 \mbox{Sector 5:} \;\;\;\; &
 L^{H}_{1}
 & = \;\; & 
 \eps
 \left(\frac{-s}{\mu^2}\right)
 \; {\bf D}^- I^H_{1010}
 & \; = \; &
 -2
 + \mathrm{subleading} + {\mathcal O}\left(\eps^3\right),
 \nonumber \\
 \mbox{Sector 10:} \;\;\;\; &
 L^{H}_{2}
 & = \;\; & 
 \eps 
 \left(\frac{-t}{\mu^2}\right)
 \; {\bf D}^- I^H_{0101}
 & \; = \; &
 -2 +2 L_t \eps - L_t^2 \eps^2
 + \mathrm{subleading} + {\mathcal O}\left(\eps^3\right),
 \nonumber \\
 \mbox{Sector 15:} \;\;\;\; &
 L^{H}_{3}
 & = \;\; & 
 \eps^2 
 \left(\frac{-s}{\mu^2}\right)
 \left(\frac{-t}{\mu^2}\right)
 \; I^H_{1111}
 & \; = \; &
 4 -2 L_t \eps
 + \mathrm{subleading} + {\mathcal O}\left(\eps^3\right).
\end{alignat}

\end{appendix}

{\footnotesize
\bibliography{/home/stefanw/notes/biblio}
\bibliographystyle{/home/stefanw/latex-style/h-physrev5}
}

\end{document}